# Phase Stability and Sodium-Vacancy Orderings in a NaSICON Electrode

Ziliang Wang,[1] Sunkyu Park,[2,3,4] Zeyu Deng,[1] Dany Carlier, [3,4] Jean-Noël Chotard,[2,4,#] Laurence Croguennec,[3,4] Gopalakrishnan Sai Gautam,[5] Anthony K. Cheetham,[1,6] Christian Masquelier,[2,4] Pieremanuele Canepa[1,7,*]

[1]Department of Materials Science and Engineering, National University of Singapore, 9 Engineering Drive 1, 117575, Singapore
[2]Laboratoire de Réactivité et de Chimie des Solides (LRCS), CNRS UMR 7314, Université de Picardie Jules Verne, 80039 Amiens Cedex, France
[3]CNRS, Univ. Bordeaux, Bordeaux INP, ICMCB, UMR CNRS 5026, F-33600, Pessac, France
[4]RS2E, Réseau Français sur le Stockage Electrochimique de l'Energie, FR CNRS 3459, F-80039 Amiens Cedex 1, France
[5]Department of Materials Engineering, Indian Institute of Science, Bengaluru, 560012, Karnataka, India
[6]Materials Department and Materials Research Laboratory, University of California, Santa Barbara, California 93106, USA
[7]Department of Chemical and Biomolecular Engineering, National University of Singapore, 4 Engineering Drive 4, 117585 Singapore, Singapore

Corresponding authors: [#]jean-noel.chotard@u-picardie.fr , [*]pcanepa@nus.edu.sg

**Abstract:**

We elucidate the thermodynamics of sodium (Na) intercalation into the sodium super-ionic conductor (NaSICON)-type electrode, $Na_xV_2(PO_4)_3$, for promising Na-ion batteries with high-power density. This is the first report of a computational temperature-composition phase diagram of the NaSICON-type electrode $Na_xV_2(PO_4)_3$. We identify two thermodynamically stable phases at the compositions $Na_2V_2(PO_4)_3$ and $Na_{3.5}V_2(PO_4)_3$, and their structural features are described for the first time based on our computational analysis. We unveil the crystal-structure and the electronic-structure origins of the ground-state compositions associated with specific Na/vacancy arrangements, which are driven by charge orderings on the vanadium sites. These results are significant for the optimization of high-energy and power densities electrodes for sustainable Na-ion batteries.

## Introduction

The urgent demands for the next-generation clean energy technologies put increasing pressure on researchers to develop innovative materials and devices. Due to the high energy densities and desirable electrochemical cycling performance, lithium-ion batteries (LIBs) have been extensively investigated for their use in portable devices and vehicular transportation.[1] However, the expanding LIB industry may soon be limited by the availability of lithium and specific transition metals (mostly cobalt and nickel) that are only accessible in limited geographic locations, often with socio-political instabilities and/or restrictions by government policies.[2,3] Given the widespread abundance of Na-metal resources, sodium(Na)-ion batteries (NIBs) appear to be an ideal alternative of LIBs particularly for stationary applications.[4–7]

The Natrium Super Ionic CONductor (NaSICON)[8,9] vanadium phosphate, in its $Na_3V_2(PO_4)_3$ ($N_3VP$) phase and its electrochemically-derived phases, $Na_1V_2(PO_4)_3$ ($N_1VP$) and $Na_4V_2(PO_4)_3$ ($N_4VP$), has received significant attention as a promising positive electrode material for NIBs.[10–13] Throughout this manuscript, we use $N_xVP$ to indicate $Na_xV_2(PO_4)_3$ at a Na composition, x. Several investigations have been conducted to improve the electrochemical performance of $N_xVP$ (energy density ~370 Wh.kg$^{-1}$ theoretically cycling at $N_3VP$-$N_1VP$), for example, by mixing V partially with other transition metals.[11,12,14,15] Specifically, in the composition range $1 \leq x \leq 3$, $N_xVP$ intercalates Na at a relatively high voltage (~3.4 V vs. Na/Na$^+$)[10] and displays a limited volume change upon reversible Na insertion (i.e.,. ~234.41 Å$^3$/f.u. in $N_3VP$ and ~219.50 Å$^3$/f.u. in $N_1VP$). In contrast, reversible Na (de)intercalation in layered transition metal oxides typically provides lower voltages than NaSICON electrodes, as well as volume changes that may curb their use in practical NIBs.[6,16–18]

As shown in **Figure 1(a)**, the framework of $N_xVP$ is built around a common structural motif — the "lantern unit"— consisting of two $VO_6$ octahedra sharing corners with three $PO_4$ tetrahedra moieties.[8,9,12,19] In principle, the NaSICON structure can host up to 4 Na ions to form $N_4VP$[6,20,21] with a rhombohedral ($R\bar{3}c$) structure.[22–24] In practice, the $Na^+$ extraction/insertion is limited to only 3 ions (per two vanadiums) in the range $N_1VP$-$N_4VP$. It is impossible to extract the 4th Na ion from $N_1VP$ although the V(IV)/V(V) redox couple is theoretically accessible.[11,25,26] In the $R\bar{3}c$ $N_xVP$, two distinct Na sites exist: Na(1) and Na(2). The Na(1) site is 6-coordinated and sandwiched between 2 $VO_6$ octahedra, whereas Na(2) (yellow or black-outlined circles in **Figure 1(a)**) occupies the "interstitials" formed by $PO_4$ units. In $N_4VP$, both Na(1) and Na(2) sites are fully occupied. A partial occupancy of Na in the Na(2) sites (such as in $N_3VP$, where sodium ordering occurs) can further reduce the $R\bar{3}c$ symmetry, resulting in a monoclinic structure.[23]

While these studies are certainly important, the community still does not fully understand the sheer complexity of the mechanism of Na extraction from $N_3VP$, itself (during the first charge of the battery), which must be clarified to make $N_xVP$ a successful commercial electrode material. Chotard et al. identified specific Na/vacancy orderings in $N_3VP$ at different temperatures using synchrotron X-ray diffraction.[23] Upon reversible Na (de)intercalation, such orderings may be facilitated by charge ordering on the vanadium sites, though the mechanism is far from being fully understood.[10,11,23] These studies suggest that a complex interplay of Na/vacancy orderings, together with the variation in the electronic structure of vanadium, drives the formation of specific $N_xVP$ phases during Na (de)intercalation, which is the focus of this work.

Here, we reveal the driving forces behind the phase transitions of the $N_xVP$ framework, which occur as temperature and Na composition ($1 \leq x \leq 4$) are varied, using a multiscale model based on density functional theory (DFT) and grand-canonical Monte Carlo (gcMC) simulations. We analyze the relevant Na site occupancies, intercalation voltages, and entropies of Na intercalation as a function of temperature. Importantly, our theoretical study reveals the Na/vacancy ordering in the thermodynamically stable phase, $N_2VP$, whose signature has been previously reported.[27] Notably, this is also the first instance where the compositional phase diagram of a NaSICON electrode is derived from first principles. The results of our work shed light on the complexity of Na (de)insertion in NaSICON electrodes for novel NIBs.

## Results

To understand the process of sodium extraction/intercalation in the $N_xVP$ structure, we use DFT calculations to parameterize a cluster expansion Hamiltonian, which in turn is used for energy evaluations within gcMC simulations, resulting in the estimation of thermodynamic properties as a function of temperature (and composition).[28–30] Details on these methodologies are provided in the Supplementary Information (SI).[31,32]

**Figure 1(b)** shows the DFT mixing energies of the $N_xVP$ system at 0 K, referenced to the $N_1VP$ and $N_4VP$ compositions, which elucidate the thermodynamic driving forces for Na (de)intercalation. The mixing energies in **Figure 1(b)** do not include entropy contributions. We enumerated all the different Na/vacancy configurations within $2 \times 2 \times 1$ supercells (with 168 atoms) of the primitive rhombohedral cell (2 f.u. with 21 atoms per f.u.) and sampled Na compositions with a step size of x = 0.25 (see details in SI). The envelope of points minimizing the mixing energy ——the convex hull—— comprises the stable Na/vacancy orderings (aqua hexagons in **Figure 1(b)**) at 0 K, namely, x = 1, 2, 3, and 4 in $N_xVP$. Other Na/vacancy configurations (blue markers in **Figure 1(b)**) are metastable (or unstable) and will decompose or phase separate into nearby stable phases at 0 K. Several metastable configurations are observed to be close to the convex hull, such as $N_{1.5}VP$ and $N_{3.5}VP$ (with energy ~4.4 meV/f.u. and ~3.8 meV/f.u. above the convex hull, respectively).

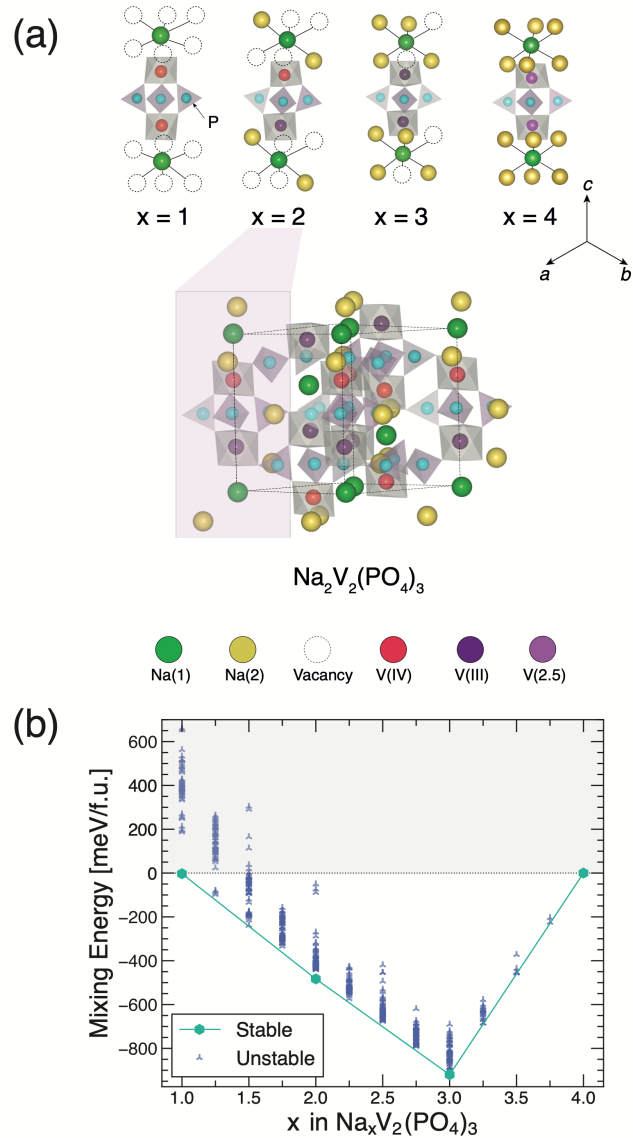

**Figure 1** Phase diagram at 0 K and stable orderings of $N_xVP$ identified by DFT. Panel (a) depicts the local environments of the stable Na/vacancy orderings of panel (b) in the format of "lantern units". Na(1) and Na(2) sites, vanadium species, i.e., V(IV), V(III), V(2.5), and P sites are identified with green, yellow, red, violet, pink, and light blue circles, respectively. A zoom out of the $N_2VP$ structure is shown emphasizing its "lantern unit" in the pink box. Vacancy $Na^+$ sites are represented by open circles. Charges on the vanadium sites are derived from the integrated magnetic moments, as calculated by DFT. Panel (b) shows the mixing energies for all Na/vacancy orderings vs. Na content (x) in $N_xVP$. The solid line (aqua) shows the convex hull envelope constructed by the most stable phases, identified by the aqua hexagons, i.e. $N_1VP$ ($R\bar{3}c$), $N_2VP$ ($P\bar{1}$), $N_3VP$ (monoclinic $Cc$), and $N_4VP$ ($R\bar{3}c$).

Our calculations indicate that the stable "end-member" compositions, namely $N_1VP$ and $N_4VP$ adopt the $R\bar{3}c$ rhombohedral space group, in agreement with existing X-ray diffraction experiments.[11] We find the monoclinic distortion of $N_3VP$ ($Cc$), which is also a global minima (~920 meV f.u.$^{-1}$) in the convex hull of **Figure 1(b)**, in agreement with experimental observation.[23] Remarkably, our DFT calculations discover a previously uncharted, thermodynamically stable Na-vacancy configuration at x = 2, i.e., $N_2VP$ (**Figure 1(a)**). In the ground state structure of $N_2VP$, the Na ions and vacancies are initialized in a monoclinic setting ($C2/c$), with geometry optimization by DFT further distorting the structure into a triclinic ($P\bar{1}$) symmetry (see further discussions below).

**Figure 1(a)** shows the "lantern units" of the computed ground states of $N_xVP$ at x = 1, 2, 3, and 4, emphasizing the Na/vacancy orderings in relation to the charge orderings on the vanadium (V) sites. The integration of the DFT-calculated spin density on each V site provides magnetic moments, which are directly related to the oxidation states of the V atoms (see **Table S4** of SI). **Table S3** of SI lists the V-O bond lengths for all the $N_xVP$ ground states, which is particularly useful for identifying V(IV) sites, as demonstrated by the existence of shorter (~1.82 Å) V(IV)-O bonds. In $N_1VP$ and $N_3VP$, only one type of V site is observed, i.e., V(IV) and V(III), respectively, which fulfils the charge-neutrality of the corresponding structures. On the other hand, the ground state configuration at $N_2VP$ exhibits "lantern units" containing one V(III) and one V(IV).

In $N_4VP$, we could not distinguish the two V sites (light violet circles in **Figure 1(a)**), as this composition shows metallic behaviour, which is discussed in the density of states (DOS) and the band structures of **Figures S1** and **S2**, respectively. The metallic behaviour of $N_4VP$ results in fractional oxidation states of V (~2.5), as extracted from the calculated magnetic moments (**Table S4**), and in agreement with previous theoretical calculations.[13] The analysis of the N$_4$VP band structure in **Figure S2** suggests that only four bands contribute actively at the Fermi energy, which indicates that the predicted metallic behaviour may be an artifact of DFT. In fact, prior hybrid DFT calculations predicted N$_4$VP as a semiconductor with a small band gap of ~0.3 eV.[13]

The projected DOS in **Figures S1(a)** to **(d)** exhibit predominant V 3$d$ states at the valence band edges and/or near the Fermi energy. As Na concentration (x) increases in N$_x$VP from 1 to 4, the V 3$d$ states are shifted to higher energies, first reducing and eventually closing the band gap at N$_4$VP. The decreasing band gap with the reduced vanadium oxidation state (and increasing Na content) in N$_x$VP also follows the trends of other compounds containing V,[33–35] where compounds with V(IV) show less tendency to undergo a semiconductor-to-metal transition near room temperature than compounds containing V(III). Hence, the qualitative trends of our band gap predictions is robust.  Notably, the appreciable increase in band gap at N$_3$VP (~1.4 eV), with respect to N$_1$VP (~0.7 eV) or N$_2$VP (~0.3 eV), is attributed to the monoclinic distortion, which stabilizes the N$_3$VP structure and shifts the occupied V 3$d$ manifolds to lower energies.[13]

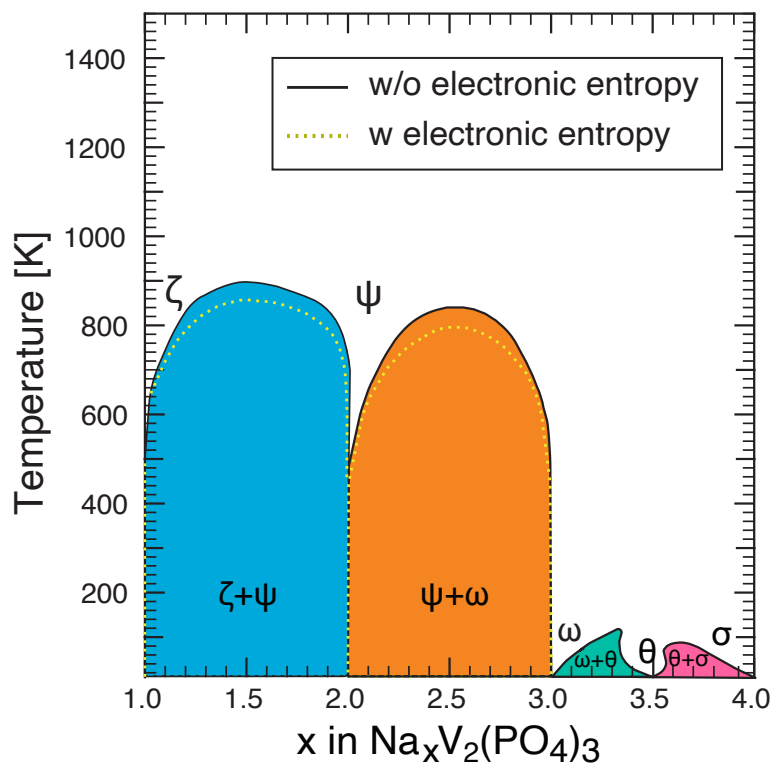

**Figure 2** Computed phase diagram of $N_xVP$ at variable Na contents (x) and temperatures (T) from gcMC simulations. Solid and dashed lines show the phase boundaries. Dashed yellow lines in the region $1 \leq x \leq 3$ represent phase boundaries obtained by introducing electronic entropy gained by mixing of the V(III) with V(IV) oxidation states.

The knowledge of the temperature vs. Na composition (T vs. x) phase diagram is also of immediate importance for understanding the extraction/intercalation of $Na^+$ from/into $N_xVP$ and for identifying commonalities in the structure-property relationships of other NaSICON electrodes or electrolytes. The temperature vs. composition phase diagram in **Figure 2** is derived from gcMC simulations, with supercells containing more than 172,032 atoms and ranged over 32,768,000-327,680,000 Monte Carlo steps for all T and x samples combined. The cluster expansion Hamiltonian, used for the energy evaluations in gcMC was fitted on DFT-calculated mixing energies of ~849 Na/vacancy configurations (see SI for more details).

**Figure 2** exhibits five single-phase regions, namely, $\zeta, \psi, \omega, \theta$ and $\sigma$, which correspond largely to Na compositions of x = 1, 2, 3, 3.5, and 4, respectively. The two-phase regions in **Figure 2** are indicated by the coloured domes, with solid black and dashed yellow lines indicating their boundaries without and with electronic entropy contributions, respectively (see discussion below). Specifically, the two-phase regions in **Figure 2** are: $\zeta + \psi$ below 890 K and at $1 < x < 2$ (light-blue dome), $\psi + \omega$ (orange dome at $2 < x < 3$, T < 840 K), $\omega + \theta$ (aqua dome at $3 < x < 3.5$, T < 120 K), and $\theta + \sigma$ (pink dome at $3.5 < x < 4$, T < 100 K). Notably, the $\zeta + \psi$, and $\psi + \omega$ two-phase regions are stable over a wide-range of temperatures (0 ~ 800 K) compared to the other two-phase regions. The $\theta$ phase detected by gcMC appears to be stable only above 10 K (not visible in **Figure 2**), and is metastable at 0 K (**Figure 1(b)**). Thus, $\theta$ decomposes to form a two-phase $\omega + \sigma$ mixture, via an eutectoid reaction at ~10 K and x = 3.5.

One remarkable finding from this phase diagram is the previously unidentified single-phase ($P\bar{1}$), $\psi$, at x = 2, which exists as a line compound up to T < 480 K. The solubility of excess Na in the $\psi$ phase increases significantly at higher temperatures (T > 480 K), eventually leading to the closing of the $\psi + \omega$ miscibility gap at ~850 K. The $\psi$ phase also exhibits Na-vacancy solubility at high temperatures (T > 630 K), resulting in the closing of the $\zeta + \psi$ two-phase region at T ~890 K. In general, we note that the $N_xVP$ phase diagram appears qualitatively similar to that of the NaSICON solid electrolyte, $Na_{1+x}Zr_2Si_xP_{3-x}O_{12}$,[19] which exhibits three distinct single-phase regions.

We assign oxidation states of V(III) and V(IV) to the vanadium atoms in $N_3VP$ and $N_1VP$, respectively, for our 0 K calculations based on integrated magnetic moments. However, at intermediate Na compositions ($1 < x < 3$), the charges V(III) and V(IV) may disorder across the available V sites, owing to thermally-activated electronic or polaronic hops,[36,37] especially at higher temperatures. Therefore, we introduce the effect of electronic entropy, $\Delta S_{\text{electronic}}$, as induced by possible disordering (or accessing multiple configurations) of V(III)/V(IV) charges via the ideal solution model (**Eq. 1**):

$$\Delta S_{\text{electronic}} = -k_{\text{B}}N[(mln(m) + (1-m)\ln{(1-m)})]  \qquad (1)$$

where, $k_{\text{B}}$ is the Boltzmann constant, $N$ is the number of vanadium sites per primitive cell (i.e., 4 sites per 2 f.u.), and $m$ is the mole fraction of V(III) oxidation states ($m = 0$ at $N_1$VP and $m = 1$ at $N_3$VP). Note that the ideal solution model provides an upper bound to the magnitude of $\Delta S_{\text{electronic}}$.

The inclusion of electronic entropy modifies the phase boundaries of the $\zeta + \psi$ and $\psi + \omega$ two-phase regions (see dashed yellow lines in **Figure 2**). Specifically, including electronic entropy lowers the critical temperatures of both the two-phase regions (i.e., the highest temperature up to which a two-phase region exists) by ~50 K. The electronic entropy also increases the Na solubility in $\psi$ phase, characterized by an increase in the stability range of the $\psi$ phase at higher Na contents for a given temperature.

We also analysed the occupancy of Na(1) and Na(2) sites, obtained from the gcMC simulations, across the Na composition range ($1 \leq x \leq 4$) at three specific temperatures: 263 K (**Figure 3(a)**), 473 K (**Figure 3(b)**) and 920 K (**Figure 3(c)**), and compared with available experimental data at 263 K and 473 K,[23] respectively. The 920 K is selected to represent the Na occupancy in the solid solution region where all two-phase regions cease to exist (**Figure 2**). For all Na compositions in $N_x$VP and at all temperatures considered, the occupation of Na(2) increases monotonically from 0 to 1 upon increasing x (see pink lines and dots in **Figure 3**). In contrast, the Na(1) site remains fully occupied at 263 K (green lines and dots) across the whole composition range. Our predictions at 263 K are in excellent agreement with the experimental value from synchrotron X-ray experiments by Chotard et al.[23] At higher temperatures (473 K and 920 K), Na(1) is fully occupied only for $1 \leq x \leq 3$, while its occupancy decreases between $N_3VP$ and $N_{3.1}VP$ and subsequently increases again from $N_{3.1}VP$ to $N_4VP$. Remarkably, at 473 K and 920 K and $x > 3.05$, both Na(1) and Na(2) sites show similar occupancies, indicating significant disorder across both Na sites, in line with the observed single-phase regions ($\omega/\theta/\sigma$) of **Figure 2**.

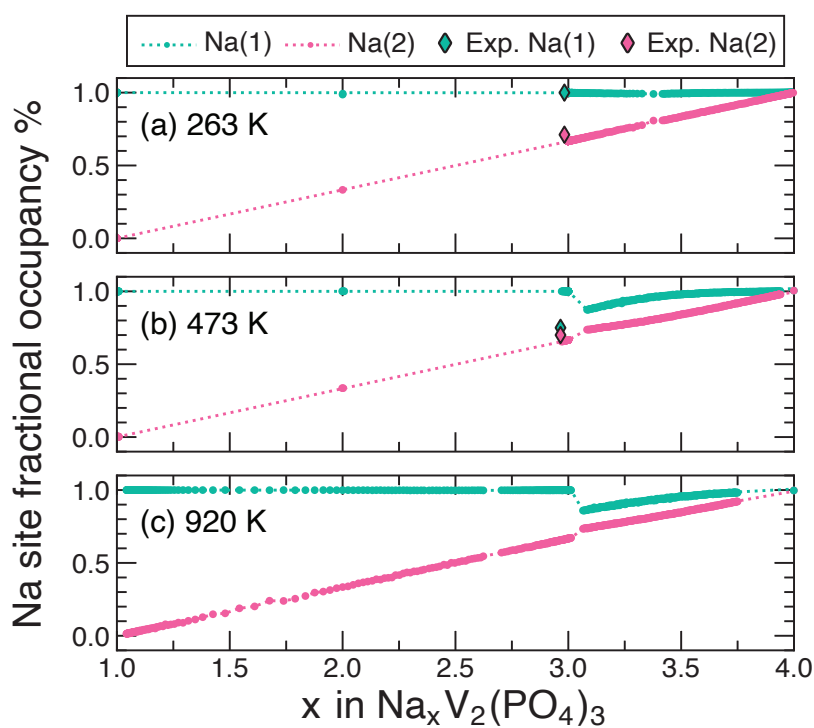

**Figure 3** Na occupancies of Na(1) sites (green) and Na(2) sites (pink) in $N_xVP$ at 263 K (panel (a)), 473K (panel (b)), and 920 K (panel (c)) vs. Na composition as extracted from gcMC simulations. Coloured dots represent data in the single-phase regions of the phase diagram of **Figure 2**. The dashed lines denote the Na(1)/Na(2) distributions within the two-phase regions and are derived by interpolation between gcMC-simulated fractional occupations of the bounding single-phase regions. The experimental occupancies of Na(1) and Na(2) at 263 K and 473 K[23] are represented by green and pink diamond, respectively.

Our simulated Na occupancy of the Na(2) site at the $N_3VP$ composition at 473 K accurately reproduces experimental data of Ref. [23]. Although we overestimate the Na(1) occupancy (~1) compared to experiment (~0.75) at $N_3VP$,[23] we predict a sharp decrease of the Na(1) occupation at x slightly above 3 (e.g., x = 3.05), which is in better agreement with the experimental occupancy of Na(1),[19] suggesting that the experimental samples intended as $N_3VP$ may be slightly off-stoichiometric.

The configurational entropy S(x) vs. Na content ($1 \leq x \leq 3$) at different temperatures is shown in **Figure S6** of SI. S(x) shows a minima at x = 1, 2, and 3, in agreement with the single-phases observed in the phase diagram (**Figure 2**). The low Na off-stoichiometry, up to temperatures as high as ~500 K at x = 1 and x = 2, reflects the low extent of configurational entropy available to these systems, as shown in **Figure S6**. Although we observe sizeable values of configurational entropy computed at high temperatures (above 800 K) in the single-phase regions ($\zeta$, $\psi$, and $\omega$), these phases are still preserved and full compositional disorder is only achieved above 900 K.

The computed $N_xVP$ voltage – composition curves at 0 K, 10 K, 300 K, and 500 K compared with the experimental voltage – composition data recorded at 298 K[11] are shown in **Figure 4**. The blue (10 K), aqua (300 K) and gold (500 K) voltage profiles are obtained from gcMC simulations, and the crimson profile is obtained directly from DFT calculations (see **Eq. 6** of the SI) at 0 K. Notably, the 0 K profile shows two voltage steps, corresponding to $\psi$ and $\omega$ single phases, and consequently three voltage plateaus, corresponding to $\zeta + \psi$ (~3.29 V vs. Na/Na+), $\psi + \omega$ (~3.24 V), and $\omega + \sigma$ (~1.89 V) two-phase regions. Generally, the computed voltage profiles at 0 K are in agreement with previous reports.[13,26,38,39]

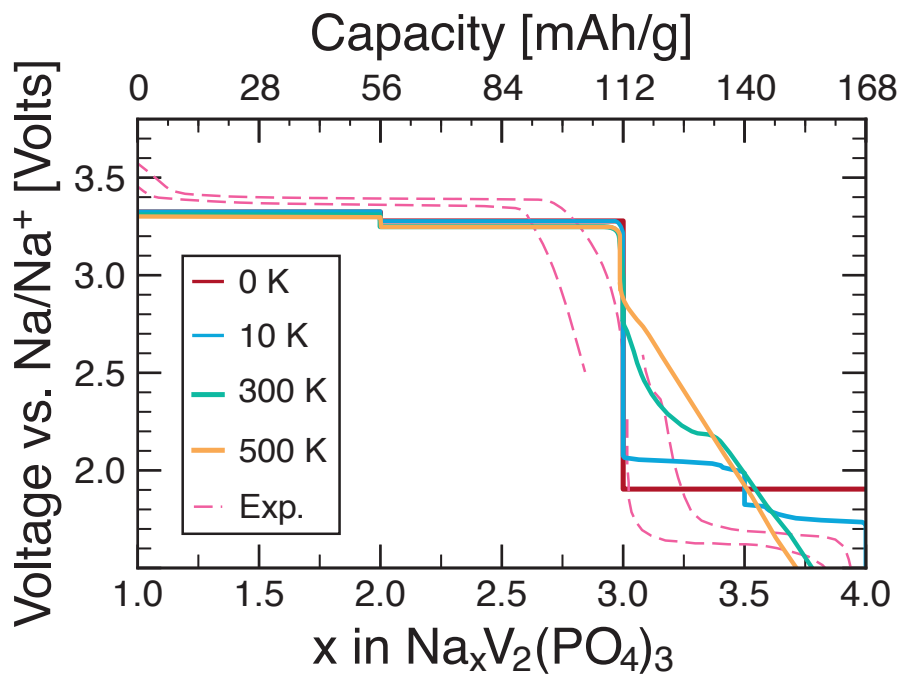

**Figure 4** Computed intercalation voltage (Volts vs. Na/Na+) curves as a function of Na content (x) in N$_x$VP. The pink dashed lines indicate the experimental voltage response of N$_x$VP at room temperature (digitized from Ref. [11] for $3 \leq x \leq 4$). Notably, the experimental profile for $1 \leq x \leq 3$ is obtained from the unpublished works of some of the authors. The crimson line shows the computed voltage curve at 0 K derived by the DFT results from **Figure 1(b)**. The blue, aqua and gold lines display the computed voltage curves at 10 K, 300 K and 500 K, respectively, generated by the gcMC results. The discharge capacities (top x-axis in the unit of mAh/g) are normalized to N$_4$VP.

Apart from a small (~50 mV) previously unreported voltage step at x = 2 (N$_2$VP), the voltage profiles for $1 \leq x \leq 3$ at different temperatures display a similar shape. Although there is good qualitative agreement between theory and experiment for the entire Na composition range, all the predicted voltage curves at the selected temperatures underestimate the average voltage across $1 \leq x \leq 3$, and overestimate the average voltage in the $3 \leq x \leq 4$ range compared to experiment.[11] Notably, the voltage curve in $3 \leq x \leq 4$ at 10 K shows a step at x = 3.5, which corresponds to the formation of $\theta$ phase via the eutectoid reaction shown in **Figure 2**. At higher temperatures (300 K and 500 K), the voltage profile for $3 \leq x \leq 4$ region exhibits a sloping feature, characteristic of the predicted single phase domain ($\omega/\theta/\sigma$) at such temperatures (**Figure 2**).

**Discussion and Conclusions**

In this letter, we have investigated the thermodynamics of Na (de)intercalation in the NaSICON electrode material, $Na_3V_2(PO_4)_3$, for Na-ion batteries. Specifically, we derived the temperature-composition phase diagram of the $N_xVP$ system and the temperature dependent voltage curves for the reversible (de)intercalation of Na-ions. Notably, we have identified two previously uncharted, thermodynamically stable phases centred around the compositions of $Na_2V_2(PO_4)_3$ and $Na_{3.5}V_2(PO_4)_3$. In addition, we have studied other important properties, such as, the ground state hull at 0 K, the Na/vacancy and charge-ordered V configurations of the 0 K ground states, and the evolution of Na(1) and Na(2) occupancies with temperature and Na content. The following paragraphs contain additional observations from the data presented in this work as well as suggestions for future work.

Unarguably, DFT-based simulations of highly correlated systems, such as $N_xVP$ (and other Na metal phosphates), present significant theoretical and computational challenges, which explain the scarce studies of such phase diagrams. Knowledge of compositional phase diagrams is however very important to accurately chart the electrochemical properties of these complex frameworks. Importantly, this is the first instance to report the composition vs. temperature phase diagram and temperature dependent voltage curves for NaSICON electrode $N_xVP$.

Three important observations can be made from the sodium occupancy data of the $N_xVP$ system: (i) The Na(1) site does not appear electrochemically active at room temperature (and below), for $1 \leq x \leq 3$, which indicates that the $Na^+$ (de)intercalation is mostly driven by the Na(2) sites; (ii) At higher temperatures (> 470 K) and higher Na contents (x > 3), both Na(1) and Na(2) attain nearly the same fractional occupations, which clearly indicates that the Na(1) site is indeed electrochemically active in these circumstances; and (iii) At x = 1, all Na(2) sites are empty and further Na removal will require the extraction of Na from Na(1) sites. With respect to observations (i) and (ii), we suggest that, at high temperature and x, $Na^+$ will be transferred between Na(1) and Na(2), which might kinetically facilitate the reversible Na (de)intercalation (at least for x > 3). Also, we expect that any electronic entropy arising from charge disordering on the vanadium sites in $1 \leq x \leq 3$ will promote disordering among the Na(1) and Na(2) sites. For observation (iii), we note that in 1992, Gopalakrishnan claimed chemical extraction of the last $Na^+$ from $N_1VP$, but this result has not been yet successfully reproduced. [11–13,39] Our data indicates that, at x = 1, all Na ions only occupy Na(1) sites, and their extraction may affect the structural integrity of $N_1VP$, thus suppressing the chemical or electrochemical extraction of the last $Na^+$. Indeed, Na migration between Na(1) sites at $N_1VP$ is hindered by high migration barriers (~755 meV, computed previously),[38] which usually indicate highly stable arrangement of atoms. Furthermore, the thermodynamic instability of the $V_2(PO_4)_3$ framework with respect decomposition into $VPO_5 + VP_2O_7$ can also account for the lack of reproduceable extraction of the last $Na^+$ from $N_1VP$.[13]

During the construction of the $N_xVP$ phase diagram, we isolated two novel low-temperature stable Na/vacancy orderings at the compositions $N_2VP$ and $N_{3.5}VP$. Other low energy structures are visible from an analysis of the convex hull of **Figure 1**, for example $N_{1.5}VP$ and $N_{3.25}VP$, but their structures are not stabilized by configurational entropy at increasing temperatures (see **Figure 2**). Notably, the signature of $N_2VP$ has been reported through *operando* X-ray diffraction by Zakharkin et al.,[27] but its structural features were never identified and reported. In addition, our gcMC predicts a new stable phase, $N_{3.5}VP$, that exists as the θ phase from 10 to ~120 K, after which it merges with the ω and σ phases into a larger monophasic region (**Figure 2**).

The ground state triclinic structure of N$_2$VP (**Figures 2** and **4**), which exists as a line compound from 0 K to ~480 K, displays a specific Na/vacancy ordering at low temperatures. At the structural level, the N$_2$VP ordering is achieved by the cooperative organization of the highly mobile Na-ions (occupying Na(2) sites) near V(III) sites and away from more positive V(IV) sites. The stability of the stoichiometric N$_2$VP ordering (i.e., the solubility of excess Na or Na-vacancies) also depends upon the adaptability of the vanadium oxidation states by inter-valence electron transfer, which can be quantified by the impact of electronic entropy on the phase boundary of the $\psi$ phase. Notably, we find that the inclusion of electronic entropy in our model only marginally modifies the N$_x$VP phase boundaries (**Figure 2**), in contrast to other phosphate electrode materials (e.g., LiFePO$_4$), where electron-hole localization affects the features of the phase diagram significantly.[41] Thus, we can conclude that charge ordering on the vanadium sites (as shown in **Figure 5**) is more likely to drive specific Na/vacancy arrangements, especially at low temperatures. Since Na ions are expected to be highly mobile in NaSICON structures (as indicated in **Figure 5**),[9] the disorder in the Na/vacancy sites is the main contributor to the configurational entropy of the $\psi$ phase at higher temperatures and not any charge-disorder on the V sites, thus restricting the extent of Na off-stoichiometry exhibited by the $\psi$ phase up to T ~ 480 K.

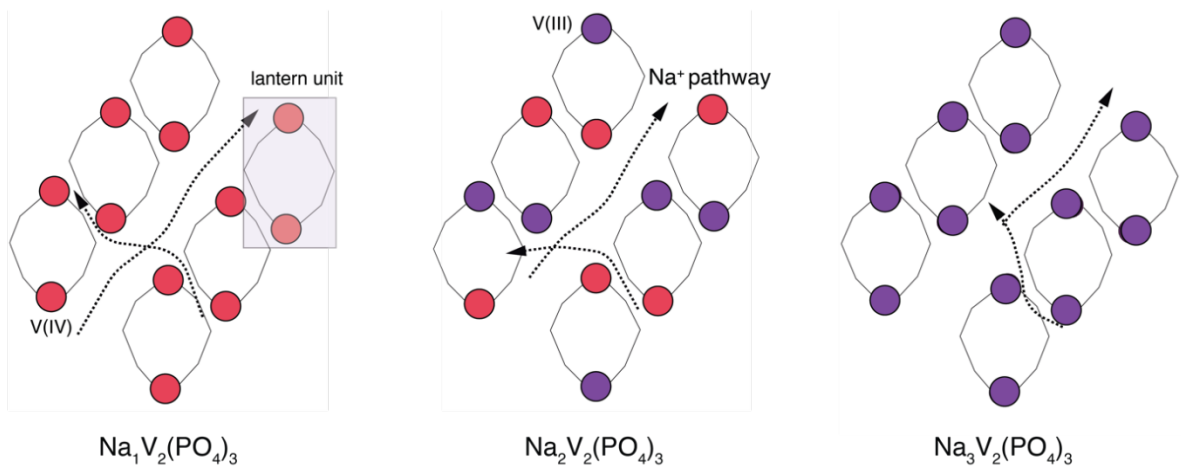

Na$_1$V$_2$(PO$_4$)$_3$  Na$_2$V$_2$(PO$_4$)$_3$  Na$_3$V$_2$(PO$_4$)$_3$

**Figure 5** Two-dimensional schematic of V charge orderings at different ground states, namely $N_1$VP, $N_2$VP and $N_3$VP. V(IV) and V(III) sites are represented by red and purple circles, respectively. The typical "lantern unit" is highlighted by the violet box. Na-ion pathways are shown with dashed arrow lines. The 3D arrangement of the "lantern" is not conveyed by this cartoon.

There are several reasons why $N_2$VP might be difficult to be observed in practice. First, we speculate that $N_2$VP might be difficult to isolate upon electrochemical (de)intercalation of Na. For example, the voltage step corresponding to $N_2$VP formation (~50 mV at x =2 in **Figure 4**) is similar to the magnitude of polarization that is typically observed in experiments. Standard electrochemical measurements may "bypass" the voltage signature of the stable $N_2$VP ordering. Second, $N_2$VP is only marginally stable with respect to disproportionation into $N_1$VP and $N_3$VP, as can be seen from **Figures 1(b)** and **4**, which may increase the difficulty of a direct solid-state synthesis of $N_2$VP from other precursors. Third, it may be simply too difficult to recognize the monoclinic or triclinic ($C2/c$ or $P\bar{1}$ according to our DFT relaxations) distorted $N_2$VP structures in the presence of the dominant rhombohedral $N_1$VP and/or $N_3$VP phases. In this context, specialized synthesis routes and/or "*operando*" experiments can be carried out to verify the existence of the $N_2$VP ordering. Previous computational studies, with different exchange-correlation functionals, had indeed assigned $N_2$VP as a slightly metastable phase (~2 meV/atom above the convex hull), which might give an indication of the error bar in our computed data.[13]

The experimental voltage curve of **Figure 4** in the region $3 \leq x \leq 3.5$ shows an appreciable sloping behaviour followed by a voltage-composition plateau in the $3.5 \leq x \leq 4$ region, which is partially captured in our calculated voltage profiles at 300 K or 500 K.[11] It is not clear whether this sloping region is due to polarization effects (during either charging or discharging) encountered in the experiment, or due to solid solution behaviour, as evidenced by our predictions in **Figures 2** and **4**. It also appears that the experiment could not achieve full Na insertion to reach $N_4VP$, which may be facilitated by higher temperature experiments, and requires further theoretical and experimental investigations.

The knowledge gained from our work can guide the interpretation of future experiments on $N_xVP$ and other high-energy density NaSICON electrodes for the development of optimized Na-ion batteries.

## Acknowledgments

P.C., C.M., A.K.C., and J.-N. C. are grateful to the ANR-NRF NRF2019-NRF-ANR073 Na-MASTER. P.C. acknowledges funding from the National Research Foundation under his NRF Fellowship NRFF12-2020-0012. L. C., D. C. and C. M. acknowledge the ANRT and TIAMAT for the funding of S. P.'s PhD thesis as well as the financial support from the Région Nouvelle Aquitaine and from the French National Research Agency (STORE-EX Labex Project ANR-10-LABX-76-01). The computational work was performed using resources of the National Supercomputing Centre, Singapore (https://www.nscc.sg).

## References

(1)  Goodenough, J. B.; Park, K.-S. The Li-Ion Rechargeable Battery: A Perspective. *J. Am. Chem. Soc.* **2013**, *135* (4), 1167–1176. https://doi.org/10.1021/ja3091438.

(2)  Olivetti, E. A.; Ceder, G.; Gaustad, G. G.; Fu, X. Lithium-Ion Battery Supply Chain Considerations: Analysis of Potential Bottlenecks in Critical Metals. *Joule* **2017**, *1* (2), 229–243. https://doi.org/10.1016/j.joule.2017.08.019.

(3)  Tarascon, J.-M. Is Lithium the New Gold? *Nature Chem* **2010**, *2* (6), 510–510. https://doi.org/10.1038/nchem.680.

(4)  Kim, S.-W.; Seo, D.-H.; Ma, X.; Ceder, G.; Kang, K. Electrode Materials for Rechargeable Sodium-Ion Batteries: Potential Alternatives to Current Lithium-Ion Batteries. *Adv. Energy Mater.* **2012**, *2* (7), 710–721. https://doi.org/10.1002/aenm.201200026.

(5)  Palomares, V.; Serras, P.; Villaluenga, I.; Hueso, K. B.; Carretero-González, J.; Rojo, T. Na-Ion Batteries, Recent Advances and Present Challenges to Become Low Cost Energy Storage Systems. *Energy Environ. Sci.* **2012**, *5* (3), 5884. https://doi.org/10.1039/c2ee02781j.

(6)  Yabuuchi, N.; Kubota, K.; Dahbi, M.; Komaba, S. Research Development on Sodium-Ion Batteries. *Chem. Rev.* **2014**, *114* (23), 11636–11682. https://doi.org/10.1021/cr500192f.

(7)  Kaufman, J. L.; Vinckevičiūtė, J.; Krishna Kolli, S.; Gabriel Goiri, J.; Van der Ven, A. Understanding Intercalation Compounds for Sodium-Ion Batteries and Beyond. *Phil. Trans. R. Soc. A.* **2019**, *377* (2152), 20190020. https://doi.org/10.1098/rsta.2019.0020.

(8)  Hong, H. Y.-P. Crystal Structures and Crystal Chemistry in the System Na1+xZr2SixP3–xO12. *Materials Research Bulletin* **1976**, *11* (2), 173–182. https://doi.org/10.1016/0025-5408(76)90073-8.

(9)  Goodenough, J. B.; Hong, H. Y.-P.; Kafalas, J. A. Fast Na+-Ion Transport in Skeleton Structures. *Materials Research Bulletin* **1976**, *11* (2), 203–220. https://doi.org/10.1016/0025-5408(76)90077-5.

(10) Saravanan, K.; Mason, C. W.; Rudola, A.; Wong, K. H.; Balaya, P. The First Report on Excellent Cycling Stability and Superior Rate Capability of $Na_3V_2(PO_4)_3$ for Sodium Ion Batteries. *Adv. Energy Mater.* **2013**, *3* (4), 444–450. https://doi.org/10.1002/aenm.201200803.

(11) Lalère, F.; Seznec, V.; Courty, M.; David, R.; Chotard, J. N.; Masquelier, C. Improving the Energy Density of $Na_3V_2(PO_4)_3$ -Based Positive Electrodes through V/Al Substitution. *J. Mater. Chem. A* **2015**, *3* (31), 16198–16205. https://doi.org/10.1039/C5TA03528G.

(12) Masquelier, C.; Croguennec, L. Polyanionic (Phosphates, Silicates, Sulfates) Frameworks as Electrode Materials for Rechargeable Li (or Na) Batteries. *Chem. Rev.* **2013**, *113* (8), 6552–6591. https://doi.org/10.1021/cr3001862.

(13) Singh, B.; Wang, Z.; Park, S.; Gautam, G. S.; Chotard, J.-N.; Croguennec, L.; Carlier, D.; Cheetham, A. K.; Masquelier, C.; Canepa, P. A Chemical Map of NaSICON Electrode Materials for Sodium-Ion Batteries. *J. Mater. Chem. A* **2021**, *9* (1), 281–292. https://doi.org/10.1039/D0TA10688G.

(14) Lalère, F.; Leriche, J. B.; Courty, M.; Boulineau, S.; Viallet, V.; Masquelier, C.; Seznec, V. An All-Solid State NASICON Sodium Battery Operating at 200 °C. *Journal of Power Sources* **2014**, *247*, 975–980. https://doi.org/10.1016/j.jpowsour.2013.09.051.

(15) Lan, T.; Ma, Q.; Tsai, C.; Tietz, F.; Guillon, O. Ionic Conductivity of $Na_3V_2P_3O_{12}$ as a Function of Electrochemical Potential and Its Impact on Battery Performance. *Batteries & Supercaps* **2021**, *4* (3), 479–484. https://doi.org/10.1002/batt.202000229.

(16) Delmas, C.; Braconnier, J.; Fouassier, C.; Hagenmuller, P. Electrochemical Intercalation of Sodium in NaxCoO2 Bronzes. *Solid State Ionics* **1981**, *3–4*, 165–169. https://doi.org/10.1016/0167-2738(81)90076-X.

(17) Toumar, A. J.; Ong, S. P.; Richards, W. D.; Dacek, S.; Ceder, G. Vacancy Ordering in O3 -Type Layered Metal Oxide Sodium-Ion Battery Cathodes. *Phys. Rev. Applied* **2015**, *4* (6), 064002. https://doi.org/10.1103/PhysRevApplied.4.064002.

(18) Hasa, I.; Mariyappan, S.; Saurel, D.; Adelhelm, P.; Koposov, A. Y.; Masquelier, C.; Croguennec, L.; Casas-Cabanas, M. Challenges of Today for Na-Based Batteries of the Future: From Materials to Cell Metrics. *Journal of Power Sources* **2021**, *482*, 228872. https://doi.org/10.1016/j.jpowsour.2020.228872.

(19) Deng, Z.; Sai Gautam, G.; Kolli, S. K.; Chotard, J.-N.; Cheetham, A. K.; Masquelier, C.; Canepa, P. Phase Behavior in Rhombohedral NaSiCON Electrolytes and Electrodes. *Chem. Mater.* **2020**, *32* (18), 7908–7920. https://doi.org/10.1021/acs.chemmater.0c02695.

(20) Lim, S. Y.; Kim, H.; Shakoor, R. A.; Jung, Y.; Choi, J. W. Electrochemical and Thermal Properties of NASICON Structured $Na_3V_2(PO_4)_3$ as a Sodium Rechargeable Battery Cathode: A Combined Experimental and Theoretical Study. *J. Electrochem. Soc.* **2012**, *159* (9), A1393–A1397. https://doi.org/10.1149/2.015209jes.

(21) Kabbour, H.; Coillot, D.; Colmont, M.; Masquelier, C.; Mentré, O. α-$Na_3M_2(PO_4)_3$ (M = Ti, Fe): Absolute Cationic Ordering in NASICON-Type Phases. *J. Am. Chem. Soc.* **2011**, *133* (31), 11900–11903. https://doi.org/10.1021/ja204321y.

(22) Yao, X.; Zhu, Z.; Li, Q.; Wang, X.; Xu, X.; Meng, J.; Ren, W.; Zhang, X.; Huang, Y.; Mai, L. 3.0 V High Energy Density Symmetric Sodium-Ion Battery: $Na_4V_2(PO_4)_3$ ‖$Na_3V_2(PO_4)_3$. *ACS Appl. Mater. Interfaces* **2018**, *10* (12), 10022–10028. https://doi.org/10.1021/acsami.7b16901.

(23) Chotard, J.-N.; Rousse, G.; David, R.; Mentré, O.; Courty, M.; Masquelier, C. Discovery of a Sodium-Ordered Form of $Na_3V_2(PO_4)_3$ below Ambient Temperature. *Chem. Mater.* **2015**, *27* (17), 5982–5987. https://doi.org/10.1021/acs.chemmater.5b02092.

(24) Jian, Z.; Yuan, C.; Han, W.; Lu, X.; Gu, L.; Xi, X.; Hu, Y.-S.; Li, H.; Chen, W.; Chen, D.; Ikuhara, Y.; Chen, L. Atomic Structure and Kinetics of NASICON $Na_xV_2(PO_4)_3$ Cathode for Sodium-Ion Batteries. *Adv. Funct. Mater.* **2014**, *24* (27), 4265–4272. https://doi.org/10.1002/adfm.201400173.

(25) Ishado, Y.; Inoishi, A.; Okada, S. Exploring Factors Limiting Three-$Na^+$ Extraction from $Na_3V_2(PO_4)_3$. *Electrochemistry* **2020**, *88* (5), 457–462. https://doi.org/10.5796/electrochemistry.20-00080.

(26) Noguchi, Y.; Kobayashi, E.; Plashnitsa, L. S.; Okada, S.; Yamaki, J. Fabrication and Performances of All Solid-State Symmetric Sodium Battery Based on NASICON-Related Compounds. *Electrochimica Acta* **2013**, *101*, 59–65. https://doi.org/10.1016/j.electacta.2012.11.038.

(27) Zakharkin, M. V.; Drozhzhin, O. A.; Ryazantsev, S. V.; Chernyshov, D.; Kirsanova, M. A.; Mikheev, I. V.; Pazhetnov, E. M.; Antipov, E. V.; Stevenson, K. J. Electrochemical Properties and Evolution of the Phase Transformation Behavior in the NASICON-Type Na3+xMnxV2-x(PO4)3 (0≤x≤1) Cathodes for Na-Ion Batteries. *Journal of Power Sources* **2020**, *470*, 228231. https://doi.org/10.1016/j.jpowsour.2020.228231.

(28) Sanchez, J. M.; Ducastelle, F.; Gratias, D. Generalized Cluster Description of Multicomponent Systems. *Physica A: Statistical Mechanics and its Applications* **1984**, *128* (1–2), 334–350. https://doi.org/10.1016/0378-4371(84)90096-7.

(29) Van der Ven, A.; Thomas, J. C.; Xu, Q.; Bhattacharya, J. Linking the Electronic Structure of Solids to Their Thermodynamic and Kinetic Properties. *Mathematics and Computers in Simulation* **2010**, *80* (7), 1393–1410. https://doi.org/10.1016/j.matcom.2009.08.008.

(30) Thomas, J. C.; Ven, A. V. der. Finite-Temperature Properties of Strongly Anharmonic and Mechanically Unstable Crystal Phases from First Principles. *Phys. Rev. B* **2013**, *88* (21), 214111. https://doi.org/10.1103/PhysRevB.88.214111.

(31) Sai Gautam, G.; Carter, E. A. Evaluating Transition Metal Oxides within DFT-SCAN and SCAN + U Frameworks for Solar Thermochemical Applications. *Phys. Rev. Materials* **2018**, *2* (9), 095401. https://doi.org/10.1103/PhysRevMaterials.2.095401.

(32) Long, O. Y.; Sai Gautam, G.; Carter, E. A. Evaluating Optimal U for 3 d Transition-Metal Oxides within the SCAN+ U Framework. *Phys. Rev. Materials* **2020**, *4* (4), 045401. https://doi.org/10.1103/PhysRevMaterials.4.045401.

(33) Feinleib, J.; Paul, W. Semiconductor-To-Metal Transition in V 2 O 3. *Phys. Rev.* **1967**, *155* (3), 841–850. https://doi.org/10.1103/PhysRev.155.841.

(34) Didier, C.; Guignard, M.; Darriet, J.; Delmas, C. O′3–Na $_x$ VO $_2$ System: A Superstructure for Na $_{1/2}$ VO $_2$. *Inorg. Chem.* **2012**, *51* (20), 11007–11016. https://doi.org/10.1021/ic301505e.

(35) Guignard, M.; Didier, C.; Darriet, J.; Bordet, P.; Elkaïm, E.; Delmas, C. P2-NaxVO2 System as Electrodes for Batteries and Electron-Correlated Materials. *Nature Mater* **2013**, *12* (1), 74–80. https://doi.org/10.1038/nmat3478.

(36) Holstein, T. Studies of Polaron Motion. *Annals of Physics* **1959**, *8* (3), 325–342. https://doi.org/10.1016/0003-4916(59)90002-8.

(37) Reticcioli, M.; Diebold, U.; Kresse, G.; Franchini, C. Small Polarons in Transition Metal Oxides. In *Handbook of Materials Modeling*; Andreoni, W., Yip, S., Eds.; Springer International Publishing: Cham, 2020; pp 1035–1073. https://doi.org/10.1007/978-3-319-44680-6_52.

(38) Ishado, Y.; Inoishi, A.; Okada, S. Exploring Factors Limiting Three-$Na^+$ Extraction from $Na_3V_2(PO_4)_3$. *Electrochemistry* **2020**, *88* (5), 457–462. https://doi.org/10.5796/electrochemistry.20-00080.

(39) Guo, X.; Wang, Z.; Deng, Z.; Wang, B.; Chen, X.; Ong, S. P. Design Principles for Aqueous Na-Ion Battery Cathodes. *Chem. Mater.* **2020**, *32* (16), 6875–6885. https://doi.org/10.1021/acs.chemmater.0c01582.

(40) Gopalakrishnan, J.; Rangan, K. K. Vanadium Phosphate (V2(PO4)3): A Novel NASICO N-Type Vanadium Phosphate Synthesized by Oxidative Deintercalation of Sodium from Sodium Vanadium Phosphate (Na3V2(PO4)3). *Chem. Mater.* **1992**, *4* (4), 745–747. https://doi.org/10.1021/cm00022a001.

(41) Zhou, F.; Maxisch, T.; Ceder, G. Configurational Electronic Entropy and the Phase Diagram of Mixed-Valence Oxides: The Case of Li x FePO 4. *Phys. Rev. Lett.* **2006**, *97* (15), 155704. https://doi.org/10.1103/PhysRevLett.97.155704.

# Phase Stability and Sodium-Vacancy Orderings in a NaSICON Electrode

## ——Supporting Information——

Ziliang Wang,[1] Sunkyu Park,[2,3,4] Zeyu Deng,[1] Dany Carlier, [3,4] Jean-Noël Chotard,[2,4,#] Laurence Croguennec,[3,4] Gopalakrishnan Sai Gautam,[5] Anthony K. Cheetham,[1,6] Christian Masquelier,[2,4] Pieremanuele Canepa[1,7,*]

[1]Department of Materials Science and Engineering, National University of Singapore, 9 Engineering Drive 1, 117575, Singapore
[2]Laboratoire de Réactivité et de Chimie des Solides (LRCS), CNRS UMR 7314, Université de Picardie Jules Verne, 80039 Amiens Cedex, France
[3]CNRS, Univ. Bordeaux, Bordeaux INP, ICMCB, UMR CNRS 5026, F-33600, Pessac, France
[4]RS2E, Réseau Français sur le Stockage Electrochimique de l'Energie, FR CNRS 3459, F-80039 Amiens Cedex 1, France
[5]Department of Materials Engineering, Indian Institute of Science, Bengaluru, 560012, Karnataka, India
[6]Materials Department and Materials Research Laboratory, University of California, Santa Barbara, California 93106, USA
[7]Department of Chemical and Biomolecular Engineering, National University of Singapore, 4 Engineering Drive 4, 117585 Singapore, Singapore

Corresponding authors: [#]jean-noel.chotard@u-picardie.fr , [*]pcanepa@nus.edu.sg

# S1. Density Functional Theory Calculations of Na$_X$V$_2$(PO$_4$)$_3$

In the reminder of this document the NaSICON material Na$_X$V$_2$(PO$_4$)$_3$ will be referred as N$_X$VP.

## S1-1. Methodology

To assess the structural, thermodynamic and electronic properties of N$_X$VP, we used the Vienna *ab initio* simulation package (VASP 6.1.0),[1,2] which implements density functional theory (DFT) to solve the Schrödinger equation of many-electron systems. The projector augmented wave (PAW) potentials, specifically Na 08Apr2002 3s$^1$, V_pv 07Sep2000 3p$^6$3d$^4$4s$^1$, P 06Sep2000 2s$^2$3p$^3$, O 08Apr2002 2s$^2$2p$^4$, were used for the description of the core electrons.[3] The valence electrons were expanded in terms of orthonormal plane-waves up to an energy cutoff of 520 eV.

To describe the electronic exchange-correlation, we employed the strongly constrained and appropriately normed (SCAN) meta-generalized gradient approximation (meta-GGA) functional.[4] SCAN includes the kinetic energy density contributions, which has been shown to remove the O$_2$ overbinding problem of GGA, while providing statistically-better ground-state electronic structures.[5] We also added a Hubbard $U$ correction of 1.0 eV on all vanadium atoms to improve the localization of 3$d$ electrons (i.e., SCAN+$U$), and thus reduce the spurious electron self-interaction error.[5] In contrast to typical GGA+$U$ calculations, a smaller $U$ value is required for SCAN calculations, whose construction limits the self-interaction error.[6,7,5] All our DFT calculations were spin-polarized and we initialized all structures in a ferromagnetic arrangement. The total energy of each calculation was converged to within 10$^{-5}$ eV/cell, atomic forces within 10$^{-2}$ eV/Å and the stress within 0.29 GPa.

To explore a sufficiently large number of Na/vacancy configurations in N$_X$VP, which is required to parameterize a cluster expansion (CE) model (see **Section S2**), we enumerated all the symmetrically distinct orderings in supercells containing as many as 8 f.u. (168 atoms) and with a resolution on the composition axis of $\Delta x = 0.25$. The enumeration process was performed with the pymatgen package.[8] At each composition, we chose a maximum of 500 structures with the lowest Ewald energy,[9] based on point charges (Na = +1, V = +2.5, P = +5, O = −2), to keep our calculations computationally tractable, followed by selecting only a subset of symmetrically distinct structures. In total, we have performed DFT structure relaxations (i.e., relax the cell volume, cell shape, and ionic positions) for 849 structures, across $1 \leq x \leq 4$ in N$_X$VP, which formed our training set for the cluster expansion. The DFT total energy of the Na/vacancy orderings was integrated on a 3×3×3 Γ-centered Monkhorst-Pack[10] $k$-point mesh for all primitive structures containing 2 N$_X$VP formula units (f.u.), 42 atoms, a 1×3×3 $k$-point mesh applied to all supercells with 4 f.u. (84 atoms), and a 1×1×3 mesh in 8 f.u. (168 atoms), respectively.

## S1-2. Lattice parameters and structures of N$_X$VP ground states

**Table S1** Fractional coordinates of atoms within each ground-state ordering (N$_1$VP, N$_2$VP, N$_3$VP, N$_4$VP) of NaSICON as computed by SCAN+$U$.

| N$_1$VP | | | | N$_2$VP | | | |
|---|---|---|---|---|---|---|---|
| Na | 0.500090 | 0.500026 | 0.499881 | Na | 0.499979 | 0.500126 | 0.499948 |
| Na | 0.000194 | 0.999900 | 0.999997 | Na | 0.999966 | 0.000008 | 0.000001 |
| V | 0.643531 | 0.643161 | 0.641416 | Na | 0.613554 | 0.247955 | 0.893272 |
| V | 0.143582 | 0.141311 | 0.143227 | Na | 0.386436 | 0.752072 | 0.106755 |
| V | 0.356334 | 0.356868 | 0.358660 | V | 0.647049 | 0.640379 | 0.643888 |
| V | 0.856446 | 0.858625 | 0.856852 | V | 0.140362 | 0.144156 | 0.149926 |
| P | 0.249946 | 0.537169 | 0.962668 | V | 0.352949 | 0.359682 | 0.356101 |
| P | 0.961382 | 0.250306 | 0.537604 | V | 0.859593 | 0.855845 | 0.850079 |
| P | 0.538668 | 0.962574 | 0.249749 | P | 0.245392 | 0.545774 | 0.966538 |
| P | 0.461350 | 0.037483 | 0.750211 | P | 0.959945 | 0.247939 | 0.542380 |
| P | 0.750062 | 0.462756 | 0.037333 | P | 0.532993 | 0.956025 | 0.249744 |
| P | 0.038567 | 0.749685 | 0.462426 | P | 0.467005 | 0.043995 | 0.750221 |
| O | 0.869278 | 0.704448 | 0.495844 | P | 0.754615 | 0.454205 | 0.033502 |
| O | 0.497020 | 0.871322 | 0.704646 | P | 0.039992 | 0.752018 | 0.457630 |
| O | 0.708184 | 0.493030 | 0.870224 | O | 0.880021 | 0.698598 | 0.498178 |
| O | 0.997173 | 0.205447 | 0.370814 | O | 0.502242 | 0.854063 | 0.750450 |
| O | 0.369284 | 0.996451 | 0.204105 | O | 0.718255 | 0.456843 | 0.876644 |
| O | 0.208056 | 0.370642 | 0.992255 | O | 0.009505 | 0.179740 | 0.381755 |
| O | 0.130606 | 0.295845 | 0.504041 | O | 0.377730 | 0.966726 | 0.204707 |
| O | 0.502663 | 0.128903 | 0.295189 | O | 0.211782 | 0.363849 | 0.025359 |
| O | 0.291615 | 0.507366 | 0.129647 | O | 0.119970 | 0.301290 | 0.501728 |
| O | 0.002735 | 0.794412 | 0.629298 | O | 0.497920 | 0.145899 | 0.249503 |
| O | 0.630745 | 0.003353 | 0.796071 | O | 0.281835 | 0.543131 | 0.123362 |
| O | 0.791668 | 0.629337 | 0.007996 | O | 0.990444 | 0.820199 | 0.618236 |
| O | 0.774757 | 0.421747 | 0.556593 | O | 0.622249 | 0.033338 | 0.795252 |
| O | 0.558242 | 0.774623 | 0.419808 | O | 0.788302 | 0.636081 | 0.974697 |
| O | 0.422212 | 0.561550 | 0.775512 | O | 0.771634 | 0.409394 | 0.553705 |
| O | 0.058294 | 0.919835 | 0.274603 | O | 0.550624 | 0.787934 | 0.428807 |
| O | 0.275055 | 0.056086 | 0.921879 | O | 0.420932 | 0.548271 | 0.782437 |
| O | 0.922325 | 0.275355 | 0.061559 | O | 0.065791 | 0.914793 | 0.265303 |
| O | 0.225189 | 0.578197 | 0.443517 | O | 0.274425 | 0.079003 | 0.919167 |
| O | 0.441974 | 0.225372 | 0.580173 | O | 0.921544 | 0.275139 | 0.071014 |
| O | 0.577956 | 0.437850 | 0.224578 | O | 0.228303 | 0.590566 | 0.446302 |
| O | 0.941952 | 0.080095 | 0.725346 | O | 0.449319 | 0.212098 | 0.571148 |
| O | 0.724976 | 0.944058 | 0.078104 | O | 0.579059 | 0.451785 | 0.217589 |
| O | 0.077766 | 0.724685 | 0.938048 | O | 0.934130 | 0.085156 | 0.734713 |
| | | | | O | 0.725550 | 0.920974 | 0.080775 |
| | | | | O | 0.078466 | 0.724827 | 0.929041 |

| N3VP | | | | N4VP | | | |
|---|---|---|---|---|---|---|---|
| Na | 0.508358 | 0.520398 | 0.473032 | Na | 0.500394 | 0.499966 | 0.499615 |
| Na | 0.008026 | 0.973669 | 0.019874 | Na | 0.000011 | 0.000360 | 0.999611 |
| Na | 0.249906 | 0.889139 | 0.618883 | Na | 0.251144 | 0.882101 | 0.616106 |
| Na | 0.615045 | 0.248978 | 0.878715 | Na | 0.615031 | 0.250504 | 0.883187 |
| Na | 0.115077 | 0.378903 | 0.748881 | Na | 0.885199 | 0.618362 | 0.248126 |
| Na | 0.749860 | 0.119149 | 0.389060 | Na | 0.118404 | 0.385126 | 0.748155 |
| V | 0.642684 | 0.643899 | 0.646102 | Na | 0.750559 | 0.114990 | 0.383249 |
| V | 0.142580 | 0.146019 | 0.144093 | Na | 0.382048 | 0.751286 | 0.116048 |
| V | 0.359385 | 0.359118 | 0.349256 | V | 0.645551 | 0.644738 | 0.645072 |
| V | 0.859500 | 0.849310 | 0.858844 | V | 0.144654 | 0.145622 | 0.145100 |
| P | 0.246064 | 0.539231 | 0.952964 | V | 0.355547 | 0.354480 | 0.354726 |
| P | 0.954651 | 0.254100 | 0.540850 | V | 0.854524 | 0.855478 | 0.854738 |
| P | 0.547487 | 0.962991 | 0.251395 | P | 0.249654 | 0.549326 | 0.950317 |
| P | 0.454738 | 0.040691 | 0.754013 | P | 0.952452 | 0.249537 | 0.548361 |
| P | 0.746215 | 0.452846 | 0.039133 | P | 0.548080 | 0.951339 | 0.250952 |
| P | 0.047583 | 0.751313 | 0.462974 | P | 0.451274 | 0.048062 | 0.750963 |
| O | 0.873367 | 0.710141 | 0.518880 | P | 0.749551 | 0.452485 | 0.048344 |
| O | 0.478837 | 0.880762 | 0.707344 | P | 0.049341 | 0.749635 | 0.450339 |
| O | 0.688599 | 0.475050 | 0.886726 | O | 0.889857 | 0.696457 | 0.482446 |
| O | 0.978908 | 0.207236 | 0.380868 | O | 0.483328 | 0.886880 | 0.699668 |
| O | 0.372843 | 0.018267 | 0.211148 | O | 0.696755 | 0.485111 | 0.888237 |
| O | 0.188877 | 0.386574 | 0.975336 | O | 0.984994 | 0.196663 | 0.388369 |
| O | 0.123876 | 0.291438 | 0.509432 | O | 0.386863 | 0.983555 | 0.199592 |
| O | 0.534356 | 0.117559 | 0.299881 | O | 0.196333 | 0.389944 | 0.982449 |
| O | 0.305498 | 0.482341 | 0.115396 | O | 0.113620 | 0.299725 | 0.517762 |
| O | 0.034366 | 0.800122 | 0.617467 | O | 0.514226 | 0.109955 | 0.306850 |
| O | 0.624430 | 0.009203 | 0.790774 | O | 0.301131 | 0.517092 | 0.110977 |
| O | 0.805709 | 0.615049 | 0.982234 | O | 0.017138 | 0.801065 | 0.611011 |
| O | 0.769932 | 0.416658 | 0.567292 | O | 0.609923 | 0.014074 | 0.806910 |
| O | 0.558710 | 0.779719 | 0.418490 | O | 0.799826 | 0.613559 | 0.017798 |
| O | 0.409964 | 0.560403 | 0.765375 | O | 0.769305 | 0.409921 | 0.565998 |
| O | 0.058078 | 0.918531 | 0.279791 | O | 0.567482 | 0.767080 | 0.409541 |
| O | 0.270369 | 0.067163 | 0.916902 | O | 0.409469 | 0.566241 | 0.767342 |
| O | 0.910279 | 0.265317 | 0.060039 | O | 0.066381 | 0.909388 | 0.267325 |
| O | 0.225267 | 0.581016 | 0.441199 | O | 0.266993 | 0.067610 | 0.909476 |
| O | 0.435046 | 0.227561 | 0.587700 | O | 0.909833 | 0.269262 | 0.066069 |
| O | 0.589385 | 0.434964 | 0.229250 | O | 0.233630 | 0.590031 | 0.431957 |
| O | 0.935286 | 0.087710 | 0.727729 | O | 0.435852 | 0.230646 | 0.590919 |
| O | 0.724783 | 0.941418 | 0.080717 | O | 0.590224 | 0.432517 | 0.231611 |
| O | 0.089125 | 0.729303 | 0.935218 | O | 0.932455 | 0.090261 | 0.731669 |
| | | | | O | 0.730680 | 0.935745 | 0.090990 |
| | | | | O | 0.090126 | 0.733666 | 0.931868 |

**Table S2** Lattice constants (in Å and °), volumes (in Å³) and space groups (Spg.) of the low-temperature-stable $N_xVP$ orderings at low temperature, namely $N_1VP$, $N_2VP$, $N_3VP$, $N_{3.5}VP$ and $N_4VP$ computed with SCAN+$U$. Note that Na intercalation in fully charged $N_1VP$ forms the fully discharged phase, $N_4VP$, resulting in a volume expansion of ~9.8 %.

| Structure | $a$ | $b$ | $c$ | $\alpha$ | $\beta$ | $\gamma$ | V / f.u. | Spg. |
|---|---|---|---|---|---|---|---|---|
| $N_1VP$ | 8.473 | 8.473 | 21.169 | 90.000 | 90.000 | 120.000 | 219.499 | $R\bar{3}c$ |
| $N_2VP$ | 8.521 | 8.613 | 8.629 | 60.350 | 61.055 | 60.540 | 227.159 | $P\bar{1}$ |
| $N_3VP$ | 15.038 | 8.729 | 8.689 | 90.000 | 124.709 | 90.000 | 234.409 | $Cc$ |
| $N_{3.5}VP$ | 8.672 | 8.694 | 15.202 | 91.110 | 105.621 | 118.923 | 237.715 | $P\bar{1}$ |
| $N_4VP$ | 8.935 | 8.935 | 20.904 | 90.000 | 90.000 | 120.000 | 241.012 | $R\bar{3}c$ |

**Table S3** Computed average (Avg.) V-O bond lengths (in Å) of different $VO_6$ octahedra represented by their V oxidation states within $N_1VP$, $N_2VP$, $N_3VP$, $N_{3.5}VP$, and $N_4VP$. The bond lengths are averaged over all the specific octahedra within each $N_xVP$ ordering. In $N_1VP$ and $N_2VP$, which contain V(IV), the minimum (Min.) and maximum (Max.) V(IV)-O bond lengths are listed instead.

| Structure | Avg. V-O bond length of | | |
|---|---|---|---|
| | $V(II)O_6$ | $V(III)O_6$ | $V(IV)O_6$ |
| $N_1VP$ | — | — | Min. 1.855; Max. 1.959 |
| $N_2VP$ | — | 2.003 | Min. 1.818; Max. 2.038 |
| $N_3VP$ | — | 2.012 | — |
| $N_{3.5}VP$ | 2.057 for $V(II)/V(III)O_6$; 2.015 for $V(III)O_6$ | | |
| $N_4VP$ | 2.062 for $V(II)/V(III)O_6$ | | |

## S1-3. Electronic Structure of N$_X$VP

**Figure S1** shows the SCAN+$U$ calculated density of states (DOS) for the four thermodynamic ground-state structures.

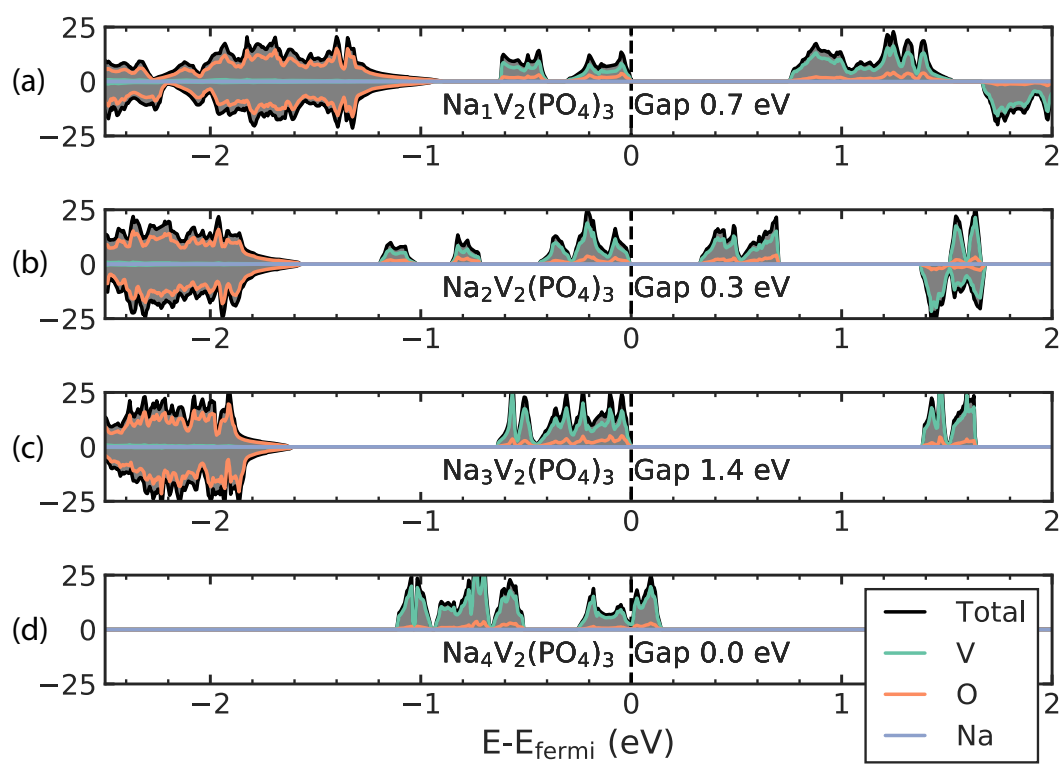

**Figure S1** The total (gray) and element-projected (green for V, orange for O, blue for Na) DOS of the most stable Na$_x$V$_2$(PO$_4$)$_3$ phases. Na contents vary from x=1 (panel (a)) to x=4 (panel (d)). The vertical dashed lines denote the Fermi energy level, and the band gap is calculated from SCAN+U based DFT.

From **Figure S1,** we deduce that the vanadium $3d$ states dominate the valence band, and the band gap for N$_1$VP, N$_2$VP, N$_3$VP and N$_4$VP is 0.7 eV, 0.3 eV, 1.4 eV, and 0 eV, respectively. Except for N$_3$VP where the gap opens due to the stabilization induced by the rhombohedral-to-monoclinic distortion, the band gap generally narrows as Na intercalation progresses from $x = 1$ to 4.

**Figure S2** shows the band structure of N$_4$VP, where only 4 bands populate the Fermi energy level.

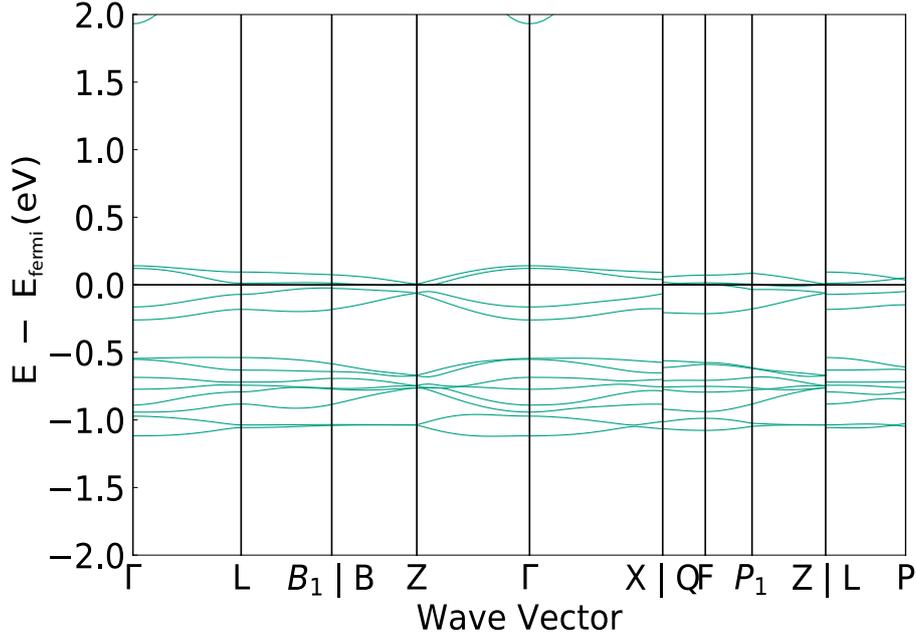

**Figure S2** The band structure of $Na_4V_2(PO_4)_3$ computed from SCAN+$U$ DFT calculation. The x axis shows the high symmetry K-point path, and the Fermi energy level is represented by the black horizontal line.

In **Table S4**, we list the computed average magnetic moments on the vanadium sites in $N_1VP$, $N_2VP$, $N_3VP$, $N_{3.5}VP$, and $N_4VP$, and assign the corresponding vanadium oxidation states.

**Table S4** The DFT-computed magnetic moments (in $\mu_B$) of vanadium sites within the NaSICON structures and assigned oxidation states.

| Compound | Oxidation States and Computed Magnetic Moments | | |
|---|---|---|---|
| | V(II) | V(III) | V(IV) |
| $N_1VP$ | — | — | $\mu_B = 1.0$ |
| $N_2VP$ | — | $\mu_B = 1.8$ | $\mu_B = 1.1$ |
| $N_3VP$ | — | $\mu_B = 1.9$ | — |
| $N_{3.5}VP$ | $\mu_B = 1.9\text{-}2.3$ | | — |
| $N_4VP$ | $\mu_B = {\sim}2.3$ | | — |

Our DFT data demonstrates that the intercalation of Na ions into the $N_1VP$ is realized by the reduction of V(IV) to V(III) to form $N_3VP$. Specifically, $N_1VP$ and $N_3VP$ are identified by a single vanadium oxidation state, namely V(IV) in $N_1VP$ and V(III) in $N_3VP$, while the charge ordering on vanadium sites in $N_2VP$ is clearly indicated by the distinct magnetic moments exhibited by V(IV) and V(III) ions. Further $Na^+$ intercalation into $N_3VP$ gives rise to $N_{3.5}VP$ and $N_4VP$, accompanied by the appearance of a fractional V oxidation state of 2.5, due the metallic transformation (see **Figures S1** and **S2**) and consequent delocalization of 3$d$ electrons on vanadium sites. Thus, we obtained average magnetic moments ranging from 1.9 $\mu_B$ to 2.3 $\mu_B$ (per vanadium), which are represented by the mixed states of V(III) and V(II).

# S2. Cluster Expansion model

## S2-1. General Theory of Cluster Expansion

We developed a CE Hamiltonian to parameterize the mixing energies ($E_{mixing}(\sigma)$ in **Eq. 7**) calculated from DFT (see **Section S1**) of various Na/vacancy orderings. The fitting of the CE was performed using the cluster assisted statistical mechanics (CASM) package.[11–14] The CE Hamiltonian was mapped onto a fixed prototypical structure, which we chose to be $N_4VP$ (see **Section S2-2**), and we wrote the CE as a truncated summation of effective cluster interactions (ECIs) composed of pair, triplet, and quadruplet clusters according to **Eq. 1**.

$$E_{mixing}(\sigma) = \sum_{\alpha} J_{\alpha} \Phi_{\alpha}(\sigma)$$
$$= \sum_{\alpha} J_{\alpha} m_{\alpha} \prod_{i \in \beta} (\sigma_i) \qquad (1)$$

where $E_{mixing}(\sigma)$ is the mixing energy as a function of Na/vacancy ordering ($\sigma$). Each term in the sum is written by the product of the ECI ($J_{\alpha}$) of cluster $\alpha$ and its cluster function $\left(\Phi_{\alpha}(\sigma)\right)$, which incorporates the multiplicity of the cluster ($m_{\alpha}$) and the correlation matrix ($\prod(\sigma)$) averaged over all clusters $\beta$ that are symmetrically equivalent to $\alpha$. Based on the Chebyshev definition, each Na site occupied by Na ion assumes $\sigma_i = -1$ and each vacancy assumes $\sigma_i = +1$. $\Phi_{\alpha}$ was generated within a radius of 10, 6, and 5 Å for the pairs, triplets, and quadruplets, respectively.

To evaluate the accuracy and predictability of the CE against the DFT mixing energy, the root mean squared error (RMS) and the leave-one-out cross-validation scores (LOOCV) were simultaneously minimized using the compressive sensing algorithm. [15] Specifically, we used a value of $\alpha = 1 \times 10^{-4}$ to penalize the L1 norm consisting of the magnitude of all ECIs and the RMS of fitted energies.[15]

## S2-2. Cluster Expansion model topology of N$_X$VP

**Table S5** shows the atom types and coordinates of the N$_4$VP rhombohedral structure, on which the CE model is developed to map the various Na/vacancy orderings. The lattice parameters of the N$_4$VP structure ($R\bar{3}c$) are *a = b = 8.936 Å, c = 20.92 Å, α = β = 90°, γ= 120°.* Note that the coordinates listed below have not been optimized with DFT.

**Table S5** Atom sites and fractional coordinates of the model topology cell of N$_4$VP. Na(1) and Na(2) sites for Na ions to occupy are consistent with the labels indicated **in Figure 1** of the main article.

| Atom Site | Site Index and type | x | y | z |
|---|---|---|---|---|
| Na/Va | 0 Na(1) | 0.500001 | 0.500001 | 0.500001 |
| Na/Va | 1 Na(1) | 0.000001 | 0.000001 | 0.000001 |
| Na/Va | 2 Na(2) | 0.116602 | 0.749998 | 0.383415 |
| Na/Va | 3 Na(2) | 0.749998 | 0.383415 | 0.116602 |
| Na/Va | 4 Na(2) | 0.383415 | 0.116602 | 0.749998 |
| Na/Va | 5 Na(2) | 0.616602 | 0.883415 | 0.249998 |
| Na/Va | 6 Na(2) | 0.883415 | 0.249998 | 0.616602 |
| Na/Va | 7 Na(2) | 0.249998 | 0.616602 | 0.883415 |
| V | 8 | 0.353862 | 0.353862 | 0.353862 |
| V | 9 | 0.853862 | 0.853862 | 0.853862 |
| V | 10 | 0.646141 | 0.646141 | 0.646141 |
| V | 11 | 0.146141 | 0.146141 | 0.146141 |
| P | 12 | 0.450427 | 0.750001 | 0.049581 |
| P | 13 | 0.750001 | 0.049581 | 0.450427 |
| P | 14 | 0.049581 | 0.450427 | 0.750001 |
| P | 15 | 0.950427 | 0.549581 | 0.250001 |
| P | 16 | 0.549581 | 0.250001 | 0.950427 |
| P | 17 | 0.250001 | 0.950427 | 0.549581 |
| O | 18 | 0.301094 | 0.111477 | 0.517259 |
| O | 19 | 0.111477 | 0.517259 | 0.301094 |
| O | 20 | 0.517259 | 0.301094 | 0.111477 |
| O | 21 | 0.801094 | 0.017259 | 0.611477 |
| O | 22 | 0.017259 | 0.611477 | 0.801094 |
| O | 23 | 0.611477 | 0.801094 | 0.017259 |
| O | 24 | 0.698911 | 0.888527 | 0.482756 |
| O | 25 | 0.888527 | 0.482756 | 0.698911 |
| O | 26 | 0.482756 | 0.698911 | 0.888527 |
| O | 27 | 0.198911 | 0.982756 | 0.388527 |
| O | 28 | 0.982756 | 0.388527 | 0.198911 |
| O | 29 | 0.388527 | 0.198911 | 0.982756 |
| O | 30 | 0.589832 | 0.232484 | 0.432262 |
| O | 31 | 0.232484 | 0.432262 | 0.589832 |
| O | 32 | 0.432262 | 0.589832 | 0.232484 |
| O | 33 | 0.089832 | 0.932262 | 0.732485 |
| O | 34 | 0.932262 | 0.732484 | 0.089832 |
| O | 35 | 0.732484 | 0.089832 | 0.932262 |
| O | 36 | 0.41017 | 0.767523 | 0.567745 |
| O | 37 | 0.767523 | 0.567745 | 0.41017 |

| O | 38 | 0.567745 | 0.41017 | 0.767523 |
| O | 39 | 0.91017 | 0.067745 | 0.267523 |
| O | 40 | 0.067745 | 0.267523 | 0.91017 |
| O | 41 | 0.267523 | 0.91017 | 0.067745 |

## S2-3. Model fitting results

**Figure S3** plots the formation (mixing) energies vs. Na compositions generated by CE (red) and DFT (blue). The corresponding error of CE model against DFT are also shown.

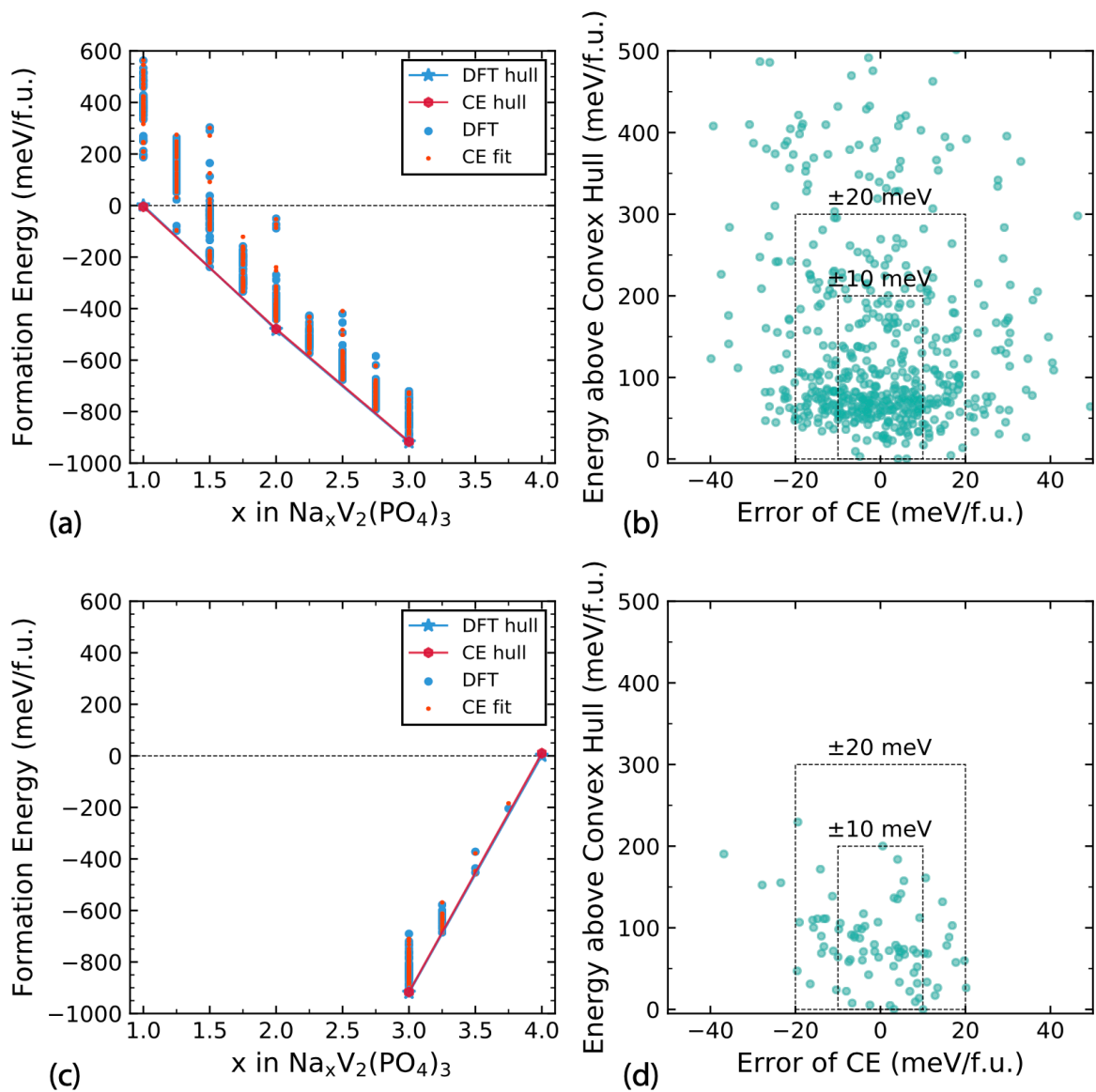

**Figure S3** DFT and CE predicted formation (mixing) energies $N_xVP$. The convex hulls with solid red line. Panel (a) shows the mixing energies of $N_xVP$ (range $1 \leq x \leq 3$) vs. Na content (x), and the corresponding error of the CE model is shown in panel (b). Panels (c) and (d) depict the mixing energies vs. Na contents (range $3 \leq x \leq 4$) and the error of the CE model, respectively. In panels (a) and (c), DFT and CE convex hulls are shown by solid blue (not visible) and red lines, where blue stars and red hexagons indicate the stable configurations forming the convex hull line. Blue and red dots depict the mixing energies of unstable configurations from DFT and the CE model respectively. In panels (b) and (d), the dashed lines denote the confidence intervals (of $\pm 10$ and $\pm 20$ meV/f.u.) of the CE models.

Separate CE models are fitted in the Na composition range $1 \leq x \leq 3$ (**Figure S3**(a)) and $3 \leq x \leq 4$ (**Figure S3**(c)), respectively, resulting in two sets of ECIs. The separation of the CE fitting is due to the differences between the electronic structure of the Na-poor (semiconducting) and Na-rich (metallic) regions of $N_x VP$ (see **Section S1-3**). Based on panels (a) and (c) of **Figure S3**, our CE models well reproduced the DFT-calculated ground state configurations, i.e., $N_1 VP$, $N_2 VP$, $N_3 VP$, and $N_4 VP$. Specifically, the RMS and LOOCV errors of the CE model for $1 \leq x \leq 3$ is ~14.85 meV/f.u. (~0.71 meV/atom), and ~22.4 meV/f.u. (~1.07 meV/atom), respectively. For $3 \leq x \leq 4$, the RMS and LOOCV of CE were ~11.45 meV/f.u. (~0.55 meV/atom), and ~18.9 meV/f.u. (~0.9 meV/atom), respectively.

To further quantify the accuracy of our CE models, the errors of the CE are shown in panels (b) and (d) of **Figure S3**. The plotted error values are the differences between the CE-predicted and corresponding DFT-calculated $E_{mixing}$ for $1 \leq x \leq 3$ (panel (b)), and $3 \leq x \leq 4$ (panel (d)), respectively.

## S2-4. Analysis of the Effective Cluster Interactions

**Table S6** lists the 33 distinctive effective cluster interactions (ECIs) of our CE model fitted in the composition region $N_1VP$–$N_3VP$.

**Table S6**. ECIs for CE model fitted on Na composition region $1 \leq x \leq 3$ of $N_xVP$. Site refers to the site labelled in **Table S5**. Min. and Max. show the minimum and maximum lengths of each ECI term, respectively, and multi. is the multiplicity of each cluster. The reference cell is labelled as [0, 0, 0].

| Cluster type | Index | Site | Cell | Min. (Å) | Max. (Å) | ECI (meV) | ECI/multi. (meV) |
|---|---|---|---|---|---|---|---|
| **Point tems** | 2 | 4 | [0, 0, 0] | — | — | 1655.913 | 275.986 |
| | 3 | 0 | [0, 0, 0] | — | — | 916.37 | 458.185 |
| **Pair Terms** | 4 | 4//0 | [0, 0, 0] [0, 0, 0] | 3.334 | 3.334 | 234.366 | 39.061 |
| | 5 | 4//1 | [0, 0, 0] [0, 0, 1] | 3.334 | 3.334 | 265.402 | 44.234 |
| | 6 | 4//6 | [0, 0, 0] [-1, 0, 0] | 4.498 | 4.498 | 31.386 | 5.231 |
| | 8 | 4//5 | [0, 0, 0] [0, -1, 0] | 4.812 | 4.812 | −26.142 | −4.357 |
| | 9 | 4//2 | [0, 0, 0] [0, 0, 0] | 4.922 | 4.922 | 14.415 | 2.403 |
| | 11 | 4//2 | [0, 0, 0] [1, -1, 0] | 5.674 | 5.674 | 15.998 | 2.666 |
| | 14 | 0//1 | [0, 0, 0] [1, 0, 0] | 6.227 | 6.227 | 18.582 | 3.097 |
| | 17 | 4//5 | [0, 0, 0] [0, 0, 0] | 6.668 | 6.668 | −11.809 | −1.968 |
| | 18 | 4//2 | [0, 0, 0] [0, -1, 0] | 6.992 | 6.992 | 100.595 | 16.766 |
| | 25 | 4//5 | [0, 0, 0] [-1, 0, 0] | 8.109 | 8.109 | −13.176 | −2.196 |
| | 31 | 4//4 | [0, 0, 0] [0, -1, 0] | 8.674 | 8.674 | −208.75 | −34.792 |
| | 32 | 4//4 | [0, 0, 0] [-1, 0, 0] | 8.674 | 8.674 | −206.17 | −34.362 |
| | 33 | 0//0 | [0, 0, 0] [0, 0, -1] | 8.674 | 8.674 | −87.795 | −14.632 |
| | 34 | 4//4 | [0, 0, 0] [0, 0, -1] | 8.674 | 8.674 | −40.975 | −6.829 |
| | 35 | 4//3 | [0, 0, 0] [0, -1, 0] | 8.841 | 8.841 | 14.931 | 2.489 |
| | 37 | 4//7 | [0, 0, 0] [1, -1, -1] | 8.875 | 8.875 | −10.004 | −1.667 |
| | 38 | 4//6 | [0, 0, 0] [-1, -1, 1] | 8.875 | 8.875 | −21.766 | −3.628 |
| | 39 | 4//4 | [0, 0, 0] [1, -1, 0] | 8.936 | 8.936 | 169.372 | 28.229 |
| | 43 | 4//5 | [0, 0, 0] [1, -1, 0] | 9.038 | 9.038 | −3.663 | −0.61 |
| | 46 | 4//1 | [0, 0, 0]<br>[1, 1, 0] | 9.416 | 9.416 | −41.499 | −6.917 |
| | 49 | 4//0 | [0, 0, 0]<br>[-1, -1, 1] | 9.416 | 9.416 | −29.701 | −4.95 |
| **Triplet Terms** | 51 | 4//6//1 | [0, 0, 0]<br>[-1, 0, 0]<br>[0, 0, 1] | 3.334 | 4.498 | 22.742 | 3.79 |
| | 53 | 4//0//2 | [0, 0, 0]<br>[0, 0, 0]<br>[0, 0, 0] | 3.334 | 4.922 | −19.119 | −3.187 |
| | 55 | 4//5//3 | [0, 0, 0]<br>[0, -1, 0]<br>[0, 0, 0] | 4.498 | 4.922 | −12.501 | −2.084 |
| | 57 | 4//2//1 | [0, 0, 0]<br>[0, -1, 1]<br>[0, 0, 1] | 3.333 | 4.922 | −63.981 | −10.664 |
| | 59 | 4//5//7 | [0, 0, 0]<br>[0, -1, 0]<br>[0, -1, 0] | 4.498 | 4.922 | −23.974 | −3.996 |
| | 60 | 4//2//3 | [0, 0, 0]<br>[0, -1, 1]<br>[-1, 0, 1] | 4.922 | 4.922 | −14.025 | −7.013 |
| | 61 | 4//5//6 | [0, 0, 0]<br>[0, -1, 0]<br>[-1, 0, 0] | 4.498 | 5.674 | −15.809 | −2.635 |
| | 62 | 4//5//2 | [0, 0, 0]<br>[0, -1, 0]<br>[1, -1, 0] | 4.498 | 5.674 | −13.268 | −2.211 |
| **Quadruplet Terms** | 66 | 4//0//2//3 | [0, 0, 0]<br>[0, 0, 0]<br>[0, 0, 0] | 3.334 | 4.922 | 31.171 | 15.586 |
| | 68 | 4//2//3//1 | [0, 0, 0]<br>[0, -1, 1]<br>[-1, 0, 1]<br>[0, 0, 1] | 3.334 | 4.922 | 5.385 | 2.693 |

**Table S7** reports the 20 distinctive ECIs in the CE model fitted on N₃VP–N₄VP region.

**Table S7** ECIs of CE model fitted on Na composition region 3 ≤ x ≤ 4 of N$_x$VP.

| cluster type | index | site | cell | Min. (Å) | Max. (Å) | ECI (meV) | ECI/multi. (meV) |
|---|---|---|---|---|---|---|---|
| Point Terms | 2 | 4 | [0, 0, 0] | — | — | −527.5 | −87.917 |
| | 3 | 0 | [0, 0, 0] | — | — | 136.003 | 68.001 |
| Pair Terms | 4 | 4//0 | [0, 0, 0]<br>[0, 0, 0] | 3.333 | 3.333 | 772.842 | 128.807 |
| | 6 | 4//6 | [0, 0, 0]<br>[-1, 0, 0] | 4.498 | 4.498 | 81.938 | 13.656 |
| | 7 | 4//7 | [0, 0, 0]<br>[0, -1, 0] | 4.498 | 4.498 | 108.13 | 18.022 |
| | 8 | 4//5 | [0, 0, 0]<br>[0, -1, 0] | 4.812 | 4.812 | 58.163 | 9.694 |
| | 9 | 4//2 | [0, 0, 0]<br>[0, 0, 0] | 4.922 | 4.922 | 104.309 | 17.385 |
| | 10 | 4//2 | [0, 0, 0]<br>[0, -1, 1] | 4.922 | 4.922 | 173.898 | 28.983 |
| | 11 | 4//2 | [0, 0, 0]<br>[1, -1, 0] | 5.674 | 5.674 | 74.069 | 12.345 |
| | 18 | 4//2 | [0, 0, 0]<br>[0, -1, 0] | 6.992 | 6.992 | 122.151 | 20.358 |
| | 22 | 4//0 | [0, 0, 0]<br>[-1, 0, 0] | 7.706 | 7.706 | 109.077 | 18.179 |
| | 32 | 4//4 | [0, 0, 0]<br>[-1, 0, 0] | 8.674 | 8.674 | −10.286 | −1.714 |
| | 40 | 4//4 | [0, 0, 0]<br>[0, 1, -1] | 8.936 | 8.936 | 3.172 | 0.529 |
| | 42 | 4//4 | [0, 0, 0]<br>[1, 0, -1] | 8.936 | 8.936 | 1.11 | 0.185 |
| Triplet Terms | 53 | 4//0//2 | [0, 0, 0]<br>[0, 0, 0]<br>[0, 0, 0] | 3.334 | 4.922 | −79.109 | −13.185 |
| | 60 | 4//2//3 | [0, 0, 0]<br>[0, -1, 1]<br>[-1, 0, 1] | 4.922 | 4.922 | −2.09 | −1.045 |
| | 62 | 4//5//2 | [0, 0, 0]<br>[0, -1, 0]<br>[1, -1, 0] | 4.498 | 5.674 | 13.296 | 2.216 |
| | 64 | 4//2//3 | [0, 0, 0]<br>[1, -1, 0]<br>[0, -1, 1] | 5.674 | 5.674 | −5.466 | −2.733 |
| Quadruplet Terms | 65 | 4//7//6//1 | [0, 0, 0]<br>[0, -1, 0]<br>[-1, 0, 0]<br>[0, 0, 1] | 3.334 | 4.922 | −26.638 | −4.44 |
| | 67 | 4//7//2//1 | [0, 0, 0]<br>[0, -1, 0]<br>[0, -1, 1]<br>[0, 0, 1] | 3.334 | 4.922 | −4.215 | −4.036 |

**Figure S4** plots the relevant most significant ECIs (normalized by their multiplicities) as a function of their cluster index.

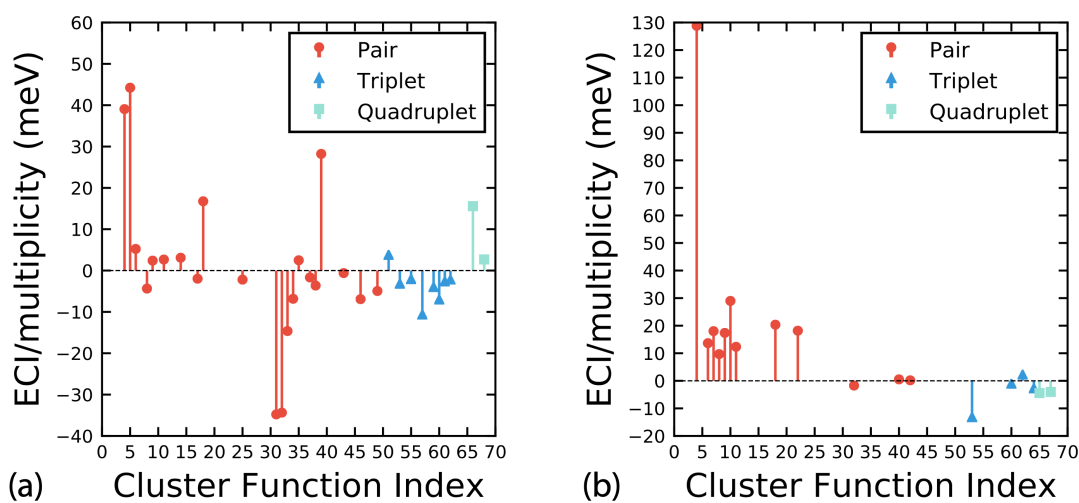

**Figure S4** Normalized ECIs vs. their cluster function index identified during the CE fitting. ECIs for the CE in the Na composition range (a) $1 \leq x \leq 3$ and (b) $3 \leq x \leq 4$.

For the CE model fitted on Na composition region $1 \leq x \leq 3$ (of **Table S6** and **Figure S4**(a)), out of the 33 clusters, 21 are pairs, 8 are triplets and 2 are quadruplets, where the pairs are the most dominant in terms of ECI/multiplicity (red dots in **Figure S4(a)**). The pairs #31 and #32 are the most stabilizing term for like-species (i.e., Na-Na or vacancy-vacancy) with ECI/multiplicity of −34.792 meV and −34.362 meV, respectively. In contrast, pairs #4 (~39.061 meV) and #5 (~44.234 meV) have a significantly stabilizing contribution for unlike-species (i.e., Na-vacancy) in our model. For the CE model fitted on Na composition range $3 \leq x \leq 4$ (of **Table S7** and **Figure S4(b)**), out of the 20 clusters, 12 are pair interactions, 4 are triplets and 2 are quadruplets, respectively. The most significant pair is #4 (~128.807 meV), which stabilizes Na-vacancy configurations.

# S3. Monte Carlo simulations

## S3-1. Methodology

We used the CASM package to perform the grand-canonical Monte Carlo (gcMC) simulations[16] on $16 \times 16 \times 16$ supercells of the primitive rhombohedral structure. The gcMC runs ranged between 32,768,000 and 327,680,000 steps and were conducted independently in three composition regions of $N_1VP$–$N_2VP$, $N_2VP$–$N_3VP$, and $N_3VP$–$N_{3.5}VP$–$N_4VP$.

## S3-2. Thermodynamic integration

The gcMC simulations were performed for $N_xVP$ system in the chemical potential ($\mu$) and temperature ($T$) space, and were then converted into ($T$, x) space to define the phase boundaries in **Figure 2**. Based on the 2 separate CE fits discussed in **Section S2**, our gcMC using CE model fitted for the composition region $N_1VP$–$N_3VP$ started from T = 10 K to 1600 K with a step of 1 K at $\mu = -4.5$, $-3.6$, and $-2.5$ eV/f.u., corresponding to $N_1VP$, $N_2VP$, and $N_3VP$, respectively. At every $T$, $\mu$ was scanned in both forward ($\mu = -4.5$ and $-3.6$ eV/f.u.) and backward ($\mu = -3.6$ and $-4.5$ eV/f.u.) directions with a step of 0.01 eV/f.u. to cover the relevant Na composition regions. Similarly, the gcMC using CE model fitted for $N_3VP$-$N_4VP$ scanned at the same temperature interval at $\mu = 6.5$, 7.2, and 9.5 eV/f.u., indicated by $N_3VP$, $N_{3.5}VP$, and $N_4VP$, respectively.

The phase boundaries were identified by the lowest envelopes of the grand-canonical potential ($\phi$) of each stable phase. $\phi$ was computed as in **Eq. 2**,

$$\phi = E - TS - \mu x \qquad (2)$$

where $E$ is the energy calculated from CE model, $T$ is the temperature, $S$ is the configurational entropy, $\mu$ is the Na chemical potential, and x represents the Na composition in $N_xVP$. To remove the numerical hysteresis for gcMC, which might cause the different voltage curves while simulating along charging/discharging process, we performed the thermodynamic integration.[17] At fixed $\mu$ and variable $T$, $\phi$ was calculated using **Eq. 3**,

$$\phi(\beta, \mu) = \frac{\beta_0}{\beta} \phi_0(\beta_0, \mu) + \frac{1}{\beta} \int_{\beta_0}^{\beta} (E - \mu x) d\beta \tag{3}$$

$$\text{with } \phi_0(\beta_0, \mu) = E - \mu x$$

where $\beta = \frac{1}{k_B T}$ , and $k_B$ is the Boltzmann's constant. The starting values $\phi_0(\beta_0, \mu)$ can be approximated as $E - \mu x$ because of the negligible entropy contribution at low temperature (i.e., $T = 10$ K).

Then at each $T$, $\phi$ was integrated by variable $\mu$ in both forward and backward directions using **Eq. 4**,

$$\phi(\beta, \mu) = \phi_0(\beta, \mu_0) - \frac{1}{\beta} \int_{\mu_0}^{\mu} x \, d\mu \tag{4}$$

$$\text{with } \phi_0(\beta, \mu_0) = \phi_{heating}(\beta, \mu_0)$$

the integration at each $\mu$ start from $\phi_{heating}(\beta, \mu_0)$, where $\mu_0 = -4.5, -3.6, -2.5$ eV/f.u. for $1 \leq x \leq 3$, and $6.5, 7.2$ and $9.5$ eV/f.u. for $3 \leq x \leq 4$, respectively.

After the thermodynamic integration the phase boundaries were found at the intersections of grand-canonical potential envelops in the (x, $T$) space, which was converted from $(\mu, T)$ space. The discontinuities in x vs $\mu$ and variations of $C_v$ vs. $\mu$ were further considered to identify the phase boundaries, as shown in **Figure S5**.

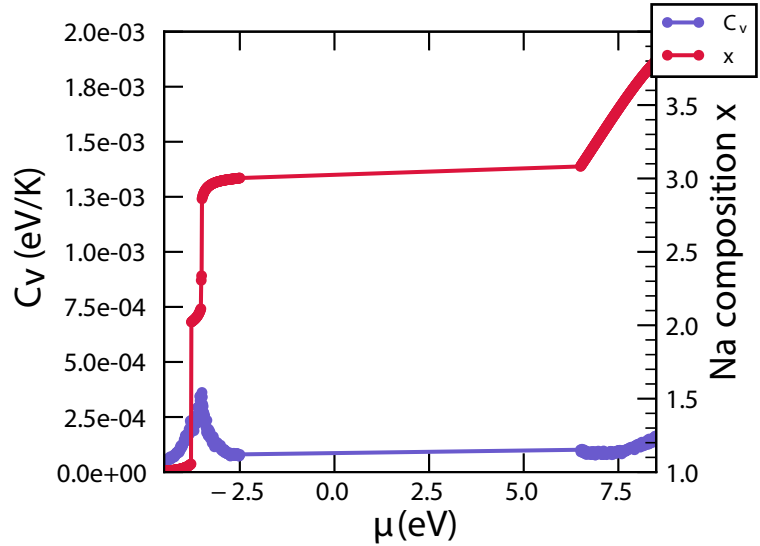

**Figure S5** Variations of normalized heat capacity $C_v$ (per f.u.) and Na composition x vs. Na chemical potential $\mu$, at $T$= 680 K, obtained from gcMC. The region of $-4.5 \leq \mu \leq -2.5$ eV/f.u. is simulated using CE model fitted for the composition range $N_1VP - N_3VP$, and the region of $\mu \geq 6.5$ eV/f.u. is simulated using CE model fitted for the composition range $N_3VP - N_4VP$. The discontinuities of x vs. $\mu$

curve (red) indicates the stable single-phases at x = 1, 2, 3, and their relevant phase boundaries. The solid solution behavior is also shown with the continuous sloping curve at $N_3VP - N_4VP$ region.

# S3-3. Configuration entropy

**Figure S6** plots the configurational entropy for Na composition region $1 \leq x \leq 3$ generated by gcMC.

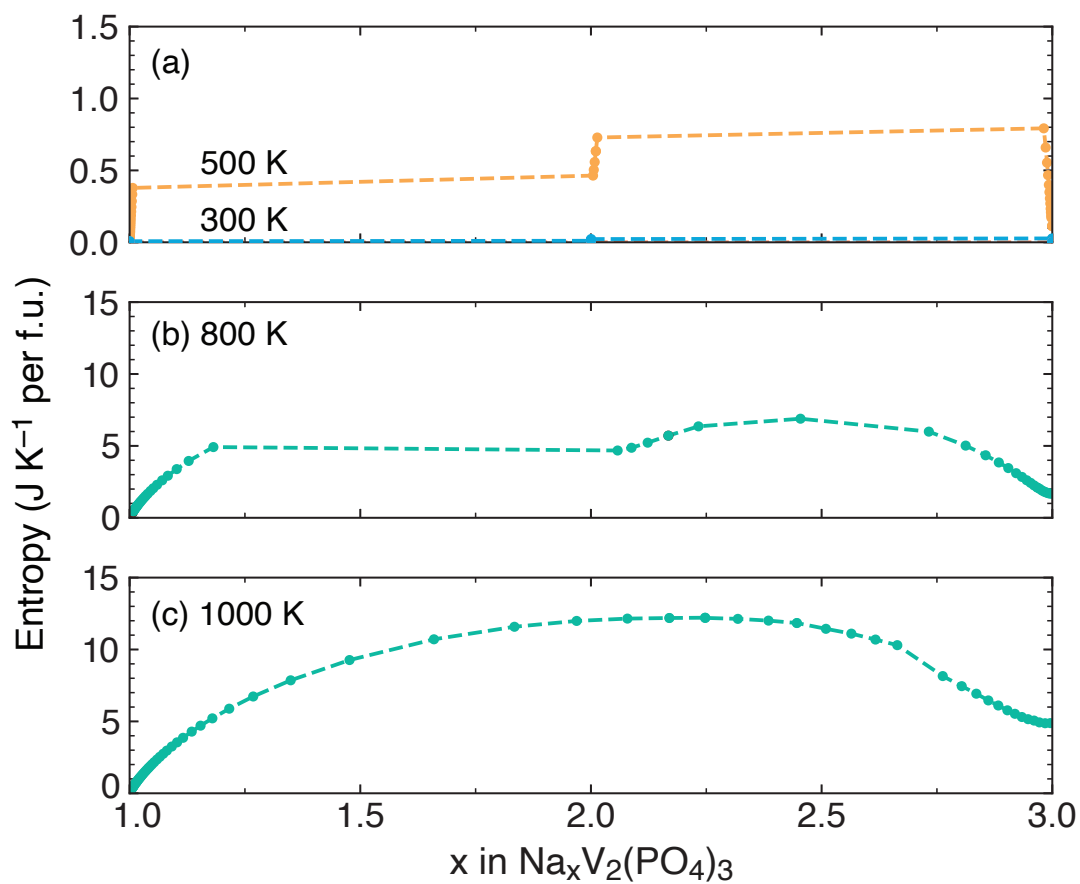

**Figure S6** Computed configurational entropy S(x) obtained from gcMC simulations as a function of Na content x in N$_x$VP and at specific temperatures 300 K, 500 K (panel(a)), 800 K (panel(b)), and 1000 K (panel(c)).

## S3-4. Computing Voltage Curves and Mixing Energies from DFT and gcMC Simulations

Reversible Na⁺ insertion/extraction into/from N$_x$VP structure is given by **Eq. 5**.

$$Na_xV_2(PO_4)_3 + yNa^+ + ye^- \xleftrightarrow{\Delta G^0} Na_{x+y}V_2(PO_4)_3 \tag{5}$$

where x indicates the initial Na content and y indicates the number of inserted Na⁺. $\Delta G^0$ is the change of Gibbs free energy for the reaction of **Eq. 5**, can be approximated from DFT total energies by ignoring the zero-point energy correction, $pV$, and entropic effects. Notably, configurational entropic effects can be included in $\Delta G^0$ using the statistical sampling from gcMC simulations.

The corresponding average intercalation voltage for the reaction of **Eq. 5** is derived based on $\Delta G^0$ using **Eq. 6**.

$$V = -\frac{\Delta G^0}{yF} \approx -\frac{E(N_{x+y}VP) - [E(N_xVP) + y\mu_{Na}]}{yF} \tag{6}$$

where $\mu_{Na}$ is the Na chemical potential (set to the bulk Na metal) and $F$ is the Faraday constant.

To evaluate the phase stability while intercalating Na⁺ into the N$_x$VP structure, we computed the mixing energies ($E_{mixing}$) at different Na compositions, using **Eq. 7**, and defined with respect to the energy of N$_1$VP and N$_4$VP end-member compositions.

$$E_{mixing}(x) = E[N_xVP] - \left(\frac{4-x}{3}\right)E[N_1VP] - \left(\frac{x-1}{3}\right)E[N_4VP] \tag{7}$$

where $E[N_xVP]$, $E[N_1VP]$, and $E[N_4VP]$ are the DFT energies of a given N$_x$VP Na/vacancy orderings, the fully discharged (N$_4$VP), and the fully charged (N$_1$VP) structures. Note, the mixing energies can be used interchangeably with formation energies in our study. The structures with the lowest mixing energies (i.e., N$_1$VP, N$_2$VP, N$_3$VP, and N$_4$VP) were used to construct the convex hull through a convex minimization at 0 K.

Additionally, the voltage curves at different temperatures are calculated from the Na chemical potential $\mu_{Na}(x)$ of gcMC, using **Eq. 8**.

$$V(x) = -\mu_{Na}(x) + \Delta\mu_{shift} \qquad (8)$$

The voltage at each Na composition is obtained by applying a chemical potential shift, $\Delta\mu_{shift}$, to $\mu_{Na}(x)$ from the gcMC. $\Delta\mu_{shift}$ measures the difference between the DFT-derived average voltage at a specific Na composition range (i.e., $1 \leq x \leq 3$, or $3 \leq x \leq 4$) and the gcMC average chemical potential $\mu_{Na}$ for the same range, referenced to the structure with the least configurational entropy (since least configurational entropy $\approx$ negligible shift in $G$ with $T$).

## References

[1]  G. Kresse, J. Furthmüller, *Comput. Mater. Sci.* **1996**, *6*, 15–50.

[2]  G. Kresse, J. Furthmüller, *Phys. Rev. B* **1996**, *54*, 11169–11186.

[3]  G. Kresse, D. Joubert, *Phys. Rev. B* **1999**, *59*, 1758–1775.

[4]  J. Sun, A. Ruzsinszky, J. P. Perdew, *Phys. Rev. Lett.* **2015**, *115*, 036402.

[5]  O. Y. Long, G. Sai Gautam, E. A. Carter, *Phys. Rev. Mater.* **2020**, *4*, 045401.

[6]  D. A. Kitchaev, H. Peng, Y. Liu, J. Sun, J. P. Perdew, G. Ceder, *Phys. Rev. B* **2016**, *93*, 045132.

[7]  G. Sai Gautam, E. A. Carter, *Phys. Rev. Mater.* **2018**, *2*, 095401.

[8]  S. P. Ong, W. D. Richards, A. Jain, G. Hautier, M. Kocher, S. Cholia, D. Gunter, V. L. Chevrier, K. A. Persson, G. Ceder, *Comput. Mater. Sci.* **2013**, *68*, 314–319.

[9]  P. P. Ewald, *Ann. Phys.* **1921**, *369*, 253–287.

[10]  J. D. Pack, H. J. Monkhorst, *Phys. Rev. B* **1977**, *16*, 1748–1749.

[11]  A. Van der Ven, J. C. Thomas, Q. Xu, J. Bhattacharya, *Math. Comput. Simul.* **2010**, *80*, 1393–1410.

[12]  B. Puchala, A. Van der Ven, *Phys. Rev. B* **2013**, *88*, 094108.

[13]  J. C. Thomas, A. V. der Ven, *Phys. Rev. B* **2013**, *88*, 214111.

[14]  CASM Developers, *Casmcode: V0.2.1*, Zenodo, **2017**.

[15]  L. J. Nelson, G. L. W. Hart, F. Zhou, V. Ozoliņš, *Phys. Rev. B* **2013**, *87*, 035125.

[16]  K. Binder, D. Heermann, L. Roelofs, A. J. Mallinckrodt, S. McKay, *Comput. Phys.* **1993**, *7*, 156.

[17]  A. van de Walle, M. Asta, *Model. Simul. Mater. Sci. Eng.* **2002**, *10*, 521–538.